\documentclass[a4paper,11pt]{article}
\pdfoutput=1 
\RequirePackage{silence}
\usepackage{jcappub} 
\usepackage[T1]{fontenc} 
\usepackage{bm}
\let\vec\bm
\usepackage{soul}

\newcommand{\T}{\mathbf{\hat{\mathcal{T}}}}

\newcommand{\ve}[1]{{\text{\bf #1}}} 
\newcommand{\vk}{\vec k}
\newcommand{\vp}{\vec p}

\newcommand{\vhn}{\hat{\vec n}}

\newcommand{\A}{\mathcal{A}}
\newcommand{\B}{\mathcal{B}}
\newcommand{\mA}{\mathcal{A}}
\newcommand{\mB}{\mathcal{B}}

\newcommand{\ikk}{\underset{\vk_{12}= \vk}{\int}}

\newcommand{\ip}{\int_{\vp}}

\newcommand{\hmpc}{\,h^{-1}\text{Mpc}}

\usepackage{booktabs}
\usepackage{float} 

\title{Fullshape power spectrum for the Symmetron modified gravity model}

\author[a,b]{Gerardo Morales-Navarrete}
\emailAdd{moralesnavarretegerardo@gmail.com}

\author[a]{Jorge L.~Cervantes-Cota}
\emailAdd{jorge.cervantes@inin.gob.mx}

\affiliation[a]{Departamento de F\'isica, Instituto Nacional de Investigaciones Nucleares,
Apartado Postal 18-1027, Col. Escand\'on, Ciudad de M\'exico,11801, M\'exico.}

\affiliation[b]{Instituto de F\'isica, Universidad Nacional Aut\'onoma de M\'exico, Apartado Postal 20-364, Ciudad de M\'exico, M\'exico.}

\abstract{ 
We make use of the perturbation theory for modified gravity models that we developed in previous works and apply it to construct the fullshape galaxy power spectrum for the Symmetron modified gravity model. First, we study the growth rate, that is a scale dependent  quantity,  and compare our results with those of the $n=1 $ Hu-Sawcki (HS) model, finding  that the Symmetron has a growth quite similar to the HS F6 in the wavenumber interval  $0.01 \leq k \leq 0.1 $ and for redshifts where Symmetron model is viable. We also propose a growth parametrization that turns to be a good approximation for the HS and Symmetron models, with a deviation less than $0.6 \%$.  
To compute the RSD multipoles we employ an expansion of the  velocity moments generating function that is suitable for general modified gravity models.  Later, we apply the fk-Perturbation Theory (fkPT) approximation to reduce the computation time of nonlinear kernels, to find the fullshape galaxy power spectrum for the Symmetron, and study the differences with HS model. The RSD multipoles of the Symmetron result similar to those of the HS F6 model. Next, we integrate this theory to an MCMC sampler and  validate our results by fitting our parameters to EZMocks to recover the parameters that bring the model to GR. We found a similar agreement in the model validation between Symmetron and F6 model, recovering the simulation cosmological parameters, and concluding that our pipeline is ready to make cosmological parameters' inference with real data. }


\begin{document} 
\maketitle
\flushbottom

\begin{section}{Introduction}

The standard model of cosmology, $\Lambda$CDM, accurately explains the anisotropies in the cosmic microwave background (CMB) radiation, as well as measurements of baryon acoustic oscillations (BAO), supernova distance-redshift relationships, among other datasets. It assumes that the way to understand the Universe on large scales is through the gravitational model based on general relativity (GR). On this framework, acceleration is hypothesized to originate from an unknown dark energy associated with the cosmological constant, but it brings along challenges such as the coincidence problem and the observed small value of its energy density compared with the value predicted by the standard particle model \cite{Clifton:2011jh,NOJIRI201159,Nojiri_2017}. This can be alleviated by alternatives, such as introducing a variable dark energy density that contributes  negatively to pressure, and hence yielding the accelerated expansion. However, this approach does not enhance our understanding of the physical origins of the theory. Furthermore, from an observational standpoint in various experiments, dark energy appeared indistinguishable from the cosmological constant when analyzing the expansion of the cosmic background. However, recent results from surveys such as DESI Year 1 \cite{DESI:2024mwx,DESI:2024kob,DESI:2024aqx} and Year 2 \cite{DESI:2025zgx,DESI:2025kuo, DESI:2025wyn}, and DES final BAO and supernova data \cite{DES:2025bxy} have shown that $\Lambda$CDM model is not necessarily the best explanation for data. Instead, models beyond the standard cosmological model such as  $\text{w}_0 \text{w}_a$CDM and wCDM have better agreement with the data. In the first case, the results of the DESI DR1 BAO data plus external datasets favor $\text{w}_0 > -1$ and $\text{w}_a < 0$ being in tension with $\Lambda$CDM at $\gtrsim 2\sigma$; also, in the second case, it occurs $\text{w} < -1$ when they consider BAO + CMB, and the SN data prefer $\text{w} > -1$. These results were further confirmed by the DESI DR2 analysis \cite{DESI:2025zgx,DESI:2025kuo, DESI:2025wyn} which concludes there is a preference for a dynamical dark energy model over $\Lambda$CDM that ranges from $2.8 - 4.2 \sigma$ depending on which supernova sample is used.  

The above considerations lead us to explore other alternative cosmological models based on Modified Gravity (MG), that may be sensible to be distinguished from the $\Lambda$CDM model through different observables, both at background and perturbative levels. MG models are of great interest when testing deviations of the standard $\Lambda$CDM cosmological model, since they are fundamental models, as opposed to parametrizations, posing a precise explanation to dark energy.  Indeed, with the observational datasets recently obtained from galaxy surveys, it is possible to test gravity on cosmological scales. This becomes even more important with data from DESI \cite{Aghamousa:2016zmz,Alam:2020jdv} and other forthcoming collaborations such as LSST \cite{Abate:2012za} and Euclid \cite{Euclid:2023bgs}. To accomplish this task, apart from the vast amount of surveys' data, there are tools such as mocks, emulators, N-body simulations, etc. \cite{Alam:2020jdv}. 
As a result of these developments, MG models can be constrained by analyzing information with different observables such as the 2-point correlation function or the power spectrum (PS), allowing us to contrast gravity models with an error at percent level accuracies \cite{Alam:2020jdv}.

In this work, we will focus on the Symmetron model \cite{Hinterbichler:2010es, Brax_2012} which has regained interest recently \cite{Ruan:2021wup,vardanyan2023modelingtestingscreeningmechanisms,Mirpoorian_2023,Fischer:2024eic}, but we also work out the  $n=1$ Hu-Sawicki $f(R)$  \cite{Hu:2007nk} that has been extensively studied in the literature \cite{Mirzatuny:2019dux,Zhao:2010qy,Ramachandra:2020lue,Rodriguez_Meza_2024} and can serve as a comparison of the different gravitational effects. Both models exhibit important features, one of them, the so-called screening mechanism, of which there are different types: Vainshtein and Chameleon are two examples. The relevance of these mechanisms is that under nonlinear effects it is possible to screen the extra degree of freedom due to the associated scalar field and recover the behavior of GR in e.g. high-density regions \cite{Aviles:2018qot}. This ensures that these gravity models comply with solar system tests where GR is well-established. On the other hand, MG models predict specific features in the expansion rate of the Universe, as well as different clustering properties, and recently the Symmetron fifth force have been explored as a particle dark matter \cite{Burrage:2016yjm,OHare:2018ayv,Burrage:2018zuj,Kading:2023hdb}. In the present study we focus on the latter properties, through the analysis of features in the power spectrum.

Understanding the large-scale structure formation of the Universe is fundamental in cosmology, especially in the context of MG theories such as the models mentioned above. While these theories can explain the origin of the accelerated expansion, modeling galaxy clustering remains challenging due to the scale dependence caused by the extra degree of freedom, apart from the other sources of non-linearity. For instance, attempts have been made to parameterize the growth rate of the structure \cite{Linder_2005, Mirzatuny:2019dux}. In \cite{Mirzatuny:2019dux} it was introduced a fitting function for the growth rate in the Hu-Sawicki $f(R)$ model that not only simplifies the calculations, but also allows an effective comparison of the results with numerical values in a wide range of scales and redshifts. Thus, as obtained in \cite{Mirzatuny:2019dux}, by achieving sub-percent accuracy in the data representation, this function becomes a valuable tool for future cosmological studies and analysis, facilitating the assessment of the influence of various parameters, such as $f_{R0}$, on the evolution of the Universe. 
Another approach is to solve the $k-$dependency model by model. This often complicates the calculations. For example, when examining the scale dependence of kernels in redshift space distortions (RSD) modeling, it is well understood that modifications are necessary for both the linear contributions (Kaiser’s formula) and the nonlinear aspects of perturbation theory (PT) in MG. Many works based on nonlinear PT in the context of the $\Lambda$CDM model have been developed for this purpose  \cite{Bernardeau_2002}. For MG, Standard Perturbation Theory (SPT) at 1-loop was developed in \cite{Koyama:2009me} and Lagrangian Perturbation Theory (LPT) in \cite{Aviles:2017aor}, and further developments on these formalisms in e.g. \cite{Aviles:2018thp,Aviles:2018saf,Aviles:2018qot,Valogiannis:2019nfz,Aviles:2019fli,Koyama:2015vza}.

On the other hand, when cosmological parameters inference is performed, there are two ways to analyze data from different surveys to extract cosmological information: BAO and fullshape analysis \footnote{aka 'full modeling', 'direct modeling'.}. The first consists of compressing data into some parameters (Alcock-Paczynski and $f \sigma_8$) and setting a PS template \cite{BOSS:2016wmc}; the second analysis \cite{DAmico:2019fhj,Ivanov:2019pdj} includes PT up to one-loop, nonlinear bias, RSD modeling and effective field theory (EFT) counterterms; with this last ingredient being useful for the validity of the analysis for large wavenumbers, until $k \sim 0.25$ $h/$Mpc.

A proposal for RSD in MG within SPT was put forward in \cite{Aviles:2020wme}, in which the nonlinear kernels are numerically solved. However, these solutions are slow for its implementation in an MCMC sampler to compute the inference of cosmological parameters from data.
To solve this problem, a fast but accurate approximation, the fkPT approach \cite{Rodriguez_Meza_2024}, has recently been developed in the context of neutrinos in the FOLPS-nu code \cite{Noriega:2022nhf} and for Hu-Sawicki $f(R)$ in the fkpt code \cite{Rodriguez_Meza_2024,Aviles:2024zlw}. The method is based on considering the scale dependence of the growth rate in the kernels, but the other $k-$dependent terms in kernels are assumed to be those of $\Lambda$CDM. Now, to correct the missing kinematics, one renormalizes  the EFT counterterms to end up with a fairly good approximation to the fullshape power spectrum; this approach has been proved to be very useful for the full shape analysis \cite{noriega2024comparingcompressedfullmodelinganalyses}. In the present work, we assume the fk-perturbation theory as implemented into the fkpt code \cite{Rodriguez_Meza_2024}, being capable of computing the redshift space galaxy power spectrum in a fraction of a second, and therefore being suitable for parameter estimation in a MCMC code.

This work is structured as follows: Section \ref{sect:basic_MGmodel} briefly reviews the most important equations  to be used here for the Hu-Sawicki-$f(R)$ and Symmetron MG models; these include the modifications in Klein-Gordon and Poisson equations at linear level. Section \ref{sect:scale_dep_growth} describes the effect of the scale dependence of the growth rate for each model and it shows plots of the growth rate as a function of the scale factor, as well as a function of the wavenumber for two $k$-bins, at the end of this Section a parametrized formula for $f(k)$ is also proposed; this parametrization is compared against Symmetron and F6 growth rate solutions. Section \ref{sect:nonlinear_MG} describes the theory for nonlinear contributions plus a infrared resummation scheme (IR), and it also explains the fk-approximation. In Section \ref{sect:multipoles} we display our results for the nonlinear contributions of the MG models using the fullshape power spectrum and the validation of the Symmetron model with EZMocks \cite{BOSS:2016wmc}. In Section \ref{sect:conclusions} we summarize our results and mention possible future work. Additionally, we include two appendices. In Appendix \ref{sect:appendix_B} we show the comparison of the multipoles for Symmetron, F6, and $\Lambda$CDM when we use $f(k,t)$ instead of $f_0(t)$ in the IR-resummation scheme for MG models and $f_0$ for $\Lambda$CDM. Appendix \ref{sect:appendix_A} presents the validation results for the HS-$f(R)$ model, similar to the procedure done in  \cite{Rodriguez_Meza_2024}.

\end{section}
\begin{section}{Modified gravity} \label{sect:basic_MGmodel}

We are interested in testing these models using the PS through a comprehensive analysis that includes PT as was mentioned above, considering contributions up to 1-loop, RSD, EFT, a bias scheme, and stochastic parameters (shot noise). Although the perturbation theory method presented throughout this work is very general and encompasses a large variety of MG theories, we will concentrate on the case of Symmetron and HS-$f(R)$ models.
\begin{subsection}{Hu-Sawicky f(R)}

The $f(R)$ gravity is defined through the action \cite{STAROBINSKY198099}
\begin{equation} \label{f_R_action}
S= \frac{1}{16 \pi G} \int d^4 x \sqrt{-g} \left( R + f(R) \right) +  S_m(\psi^i,\ve g),
\end{equation}
where $R$ is the Ricci scalar and the function $f(R)$ modifies the Einstein-Hilbert action. Further, $S_m$ is the action of matter fields $\psi^i$.

We consider the $n=1$ Hu-Sawicki $f(R)$, for short HS-$f(R)$, model given by \cite{Hu:2007nk}:
\begin{equation}
    f(R) = - \frac{c_1 R}{ c_2 R /M^2 +1}.
    \label{fR}
\end{equation}
where $c_1$ and $c_2$ are dimensionless constants to be defined next. In the following we consider a flat FRW model with perturbations in the conformal Newtonian gauge
\begin{equation}
    ds^2 = -(1+2\Phi) dt^2 + a(t)^2 (1-2\Psi)d \Vec{x}^2 .
    \label{metric}
\end{equation}
At background level, this model can be set indistinguishable from GR, by setting the constants $M^2=\Omega_m H_0^2$, $c_1/c_2= 6\Omega_{\Lambda}/\Omega_m$.
In this way, one expects deviations from GR only up to linear perturbation level. 
In the Fourier space the perturbed equations are  \cite{Koyama:2009me,Aviles:2020wme}:  
\begin{eqnarray}
    -\frac{k^2}{a^2} \Phi &=& 4\pi G \Bar{\rho} \delta + \frac{1}{2} \frac{k^2}{a^2} \delta f_R , \label{field_eqs1} \\
    3 \frac{k^2}{a^2} \delta f_R &=& 8 \pi G \Bar{\rho} \delta - M_1(k)  \delta f_R - \delta \mathcal{I} ( \delta f_R), \label{field_eqs2} \\
    \Psi - \Phi &=& \delta f_R, 
    \label{field_eqs3}
\end{eqnarray}
where $f_R \equiv df(R)/dR$ is the associated scalar field and  $\delta f_R$ its perturbation, $\Bar{\rho}$ is the background matter density (from $T^{\mu}_{\mu} = - \rho$) and $\delta$ its perturbation contrast. $M_1(k)$ is the linear contribution of the gravity model and $\delta \mathcal{I}$ encodes the nonlinear behavior that will be explained in Section \ref{sect:nonlinear_MG}.

Substituting eq. (\ref{field_eqs2}) into eq. (\ref{field_eqs1}), one obtains  
\begin{equation}
-\frac{k^2}{a^2} \Phi(\Vec{k}) = 4\pi G \overline{\rho} \mu (k,a) \delta(\Vec{k}) + S(\Vec{k}), 
\label{Phi_k}
\end{equation}
where one can identify the function:

\begin{equation}
    \mu(k,a) = 1 + \frac{k^2 / a^2}{3k^2 /a^2 + M_1(k)}.
    \label{mu_fr}
\end{equation}
that for $k=0$ one recovers GR but now at linear perturbation level and for $k\gg \sqrt{M_1 a^2 /3}$, the Newtonian effective constant becomes $G_{\rm eff} = \frac{4}{3} G$. The function $M_1$ encodes the gravitation pull at linear level, and it only depends on background quantities for this $f(R)$ model:  

\begin{eqnarray}
    M_1(a) &=& \frac{3}{2} \frac{H_0^2}{|f_{R0}|} \frac{(\Omega_{m0} a^{-3} + 4 \Omega_{\Lambda})^3}{(\Omega_{m0} + 4 \Omega_{\Lambda})^2},
    \label{M1}
\end{eqnarray}
where $f_{R0}$ is the scalar field evaluated at the background. This quantity encodes the strength of the modified gravity, whose usual values are $-10^{-4}, -10^{-5},-10^{-6}$, named F4, F5, F6, respectively. 
The second term in eq.(\ref{Phi_k}), $ S(\Vec{k})$, is defined through $\delta \mathcal{I}$ that we will further explain in Section \ref{sect:nonlinear_MG}. With the above defined quantities and field equations we will construct the growth rate in Section \ref{sect:scale_dep_growth}, but in the following, we explain the same as in this Section for the other gravity model treated in this work. 

\end{subsection}
\begin{subsection}{Symmetron}

This modified gravity model is a especial case of a general scalar-tensor theory whose  action is given by \cite{Hinterbichler:2010es, Brax_2012}:

\begin{equation}
S= \int d^4 x \sqrt{-g} \left( \frac{M_{\text{pl}}^2}{2}R - \frac{1}{2} (\partial \phi)^2 - V(\phi)  \right) +  \int d^4 x \mathcal{L}_m[\Tilde{g}],
\label{symmetron_action}
\end{equation}
here the matter fields are coupled to the metric $\Tilde{g}_{\mu \nu}$, conformally related to the Einstein frame metric by:
\begin{equation}
        \Tilde{g}_{\mu \nu} = A^2(\phi) g_{\mu \nu},
    \label{coupled_metric}
\end{equation}
where the coupling function between matter and the scalar field is
\begin{equation}
A(\phi)=1 + \frac{1}{2M^2} \phi^2 (a),
\label{coupling}
\end{equation}
with $M$ a mass scale, and from tests of gravity $M \lesssim 10^{-4}M_{pl}$ \cite{Hinterbichler:2010es, Hinterbichler:2011ca}. The background scalar field equation is given by:
\begin{equation}
    \ddot{\phi} +3H\Dot{\phi} +V_{,\phi} + \sum_i \frac{A_{,\phi}}{A^{3w_i}} (1-3w_i) \rho_i=0,
\end{equation}
where the dot over $\phi$ represents the time derivative, the sum is over the species in consideration, where we use $\rho_m \sim a^{-3}$ and $\rho_{\Lambda}=3H_0^2M_{pl}^2 (1-\Omega_m^{(0)})$. We consider the standard Symmetron model \cite{Hinterbichler:2011ca} with the symmetry-breaking potential:  
\begin{equation}
V(\phi) = \frac{\mu^4}{4 \lambda} - \frac{1}{2} \mu^2 \phi^2 + \frac{1}{4} \lambda \phi^4,
\label{potential}
\end{equation}
with $\mu$ a mass scale and $\lambda$ a dimensionless constant. In this MG model, the Universe undergoes a phase transition driven by both the scalar field potential and the matter coupling. Before the phase transition happens (at $a = a_{ssb}$) the model behaves as a $\Lambda$CDM and after that, deviations apply. The original Symmetron model, considered here, is such that the effective cosmological constant stemming from eq. (\ref{potential}) is too small when taking into account the expected particle physics phenomenology and a correct screening behavior at small scales. That is why we  introduced an extra $\rho_{\Lambda}$ in the dynamics, as done in the original model \cite{Hinterbichler:2011ca}. Therefore, the Symmetron background dynamics is quite similar to $\Lambda$CDM, but its perturbations distinguish the model.  

As aforementioned, the cosmological metric is given by eq. (\ref{metric}), that at background level has the following tracking solution valid for $a > a_{\text{ssb}}$ that should track the minimum of the effective potential:
 \begin{equation}
     \phi(a) = \phi_0 \sqrt{1- \left( \frac{a_{\text{ssb}} }{a} \right)^3 },
     \label{field}
 \end{equation}
  with $\phi_0= \mu / \sqrt{\lambda}$ and $a_{\text{ssb}}$ that corresponds to the scale factor value at which the Symmetron model undergoes its phase transition, for which we take $a_{ssb} =1/2$  ($z_{ssb}=1$); for $a<a_{\text{ssb}}$, before the transition happens, the solution is that of a  standard Einstein de-Sitter model. 
  
The Hubble equation is given by \cite{Hinterbichler:2011ca,Aviles:2010ui}:
\begin{equation}
    3 M_{pl}^{2} H^2 = \frac{1}{2} \Dot{\phi}^2 + V(\phi) + \sum_i A^{1-3w_i}(\phi) \rho_i,
    \label{Hubble}
\end{equation}
where $M_{pl}^{2} \equiv 1/8 \pi G$. Using eqs. (\ref{coupling}), (\ref{potential}), (\ref{field}) the Hubble parameter turns to be
\begin{eqnarray}
    H^2 = V_{\text{eff}} \left[ 3 M_{pl}^{2} - \frac{9}{8} \left( \frac{\phi_0}{M_{\text{pl}}} \right)^2 \left( \frac{\phi_0}{ \phi(a)} \right)^2 \left( \frac{ a_{ssb}}{a} \right)^6 \right]^{-1},
    \label{hubble}
\end{eqnarray}
where $V_{\text{eff}} = V(\phi) + A(\phi) \rho_m + A^4(\phi) \rho_{\Lambda}$ is the effective scalar field potential. The first order perturbative theory yields  the $\mu(k,a)$ for Symmetron, cf. eq. (\ref{Phi_k}) \cite{Brax_2011},  
 \begin{eqnarray}
     \mu(k,a) = 1 + \frac{2 \beta^2(a)}{A(\phi)} \frac{ k^2 }{ k^2 + a^2 m^2(a) },
     \label{mu_symmetron}
 \end{eqnarray}
where we can see for $k=0$, $\mu = 1$ recovering GR, and for $k^2 \gg a^2m^2(a)$ the Newtonian effective constant becomes $G_{\text{eff}}= \left( 1+ 2 \frac{\beta^2 (a)}{A(\phi)} \right) G$, so the fifth force is background dependent. eq. (\ref{mu_symmetron}) contains other two relevant functions for describing the Symmetron model, given by:
\begin{equation}
    m(a) = m_0 \sqrt{1- \left( \frac{a_{\text{ssb}} }{a} \right)^3} ,
    \label{mass}
\end{equation}

\begin{equation}
    \beta (a) = \beta_0 \sqrt{1- \left( \frac{a_{\text{ssb}} }{a} \right)^3},
    \label{beta}
\end{equation}
which are valid for $a>a_{\text{ssb}}$. $m$ is a scalar field mass and $\beta$ is a dimensionless function. These particular choice for $m(a)$ and $\beta(a)$ imply that $M_1$ for this MG model is a constant. In a general case, these functions, eqs. (\ref{mass}, \ref{beta}), can be generalized by having an arbitrary power law, $N$ and $M$ power, instead of $1/2$ and it is not necessary that $N$ and $M$ to be equal, for more specific details see \cite{Brax_2012}. To make connection with other works, we note that in ref. \cite{Ruan:2021wup} the authors performed Symmetron simulations, in which they also use Eqs. (\ref{mass}, \ref{beta}) but in general defined as  $\beta \equiv M_{\text{pl}} \frac{d \ln A(\phi) }{d \phi} $ and  $ m^2 \equiv \frac{d^2 V_{\text{eff}} (\phi_{\text{min}}) }{d\phi^2} $. Again, equation (\ref{mu_symmetron}) determines the growth function that we study next. Following we look at the cosmological structure formation through the linear analysis.
\end{subsection}
\end{section}

\begin{section}{Scale dependent growth rate} \label{sect:scale_dep_growth}

To see the effect of MG we can start with the differential equation for the density contrast:  

\begin{eqnarray}
    \Ddot{\delta} + 2H \Dot{\delta} - \frac{3}{2} \Omega_m H^2 \delta = 0, 
    \label{diff_eq}
\end{eqnarray}
where $\Omega_m = \rho_m(t) / \rho_c (t)$ with $\rho_c (t) = 3H^2(t) /8\pi G$. Since we will be using the code FOLPS-nu \cite{Noriega:2022nhf}, it is convenient to use the e-folds as time variable, $N= \ln(a)$, so that  $\frac{d}{dN} \equiv (~)'$. eq.(\ref{diff_eq}) can be rewritten as:
\begin{eqnarray}
    && \delta '' + \left( \frac{\Dot{H}}{H^2} + 2 \right) \delta' - \frac{3}{2} \Omega_m \delta = 0.
    \label{1era}   
\end{eqnarray}
When considering $\Lambda$CDM the term $\frac{\Dot{H}}{H^2}$ results in $ - \frac{3}{2} \Omega_m $ and the equation is given by:
 \begin{eqnarray}
 \delta '' + \left( 2 - \frac{3}{2}\Omega_m \right) \delta' - \frac{3}{2} \Omega_m \delta = 0.
 \label{delta_LCDM}
 \end{eqnarray}
However, a MG model makes different the term $\frac{\Dot{H}}{H^2}$, in general it becomes,
\begin{equation}
\frac{\Dot{H}}{H^2} = - \frac{3}{2}  \left ( 1 + w_\phi (1- \Omega_m - \Omega_\Lambda) + w_{\Lambda} \Omega_{\Lambda} \right),
\label{HoverHdot_MG}
\end{equation}
and modifies the solution of eq. (\ref{1era}) as an effect of the background alone. For $w_{\phi}=0$ and $w_\Lambda=-1$ one recovers $\Lambda$CDM. In the Symmetron case, we have that $w_\phi \ll 1$, this means that eq. (\ref{HoverHdot_MG}) is practically the $\Lambda$CDM; a similar behavior is observed for the HS-$f(R)$ model when constants are fixed in eq. (\ref{fR}) to do the background of the theory indistinguishable from $\Lambda$CDM, see discussion after eq. (\ref{metric}). 
The last term of eq. (\ref{1era}), that stems from the Poisson equation, in MG models causes the function $\mu(k,a)$ to appear as a factor, being this function responsible for  additional deviations with respect to $\Lambda$CDM.  Therefore, we obtain for the linear density contrast in MG: 
\begin{eqnarray} \label{delta_MG}
     \delta '' + \left[ 2 - \frac{3}{2}  \left ( 1 + w_\phi (1- \Omega_m - \Omega_\Lambda) + w_{\Lambda} \Omega_{\Lambda} \right)  \right] \delta' - \frac{3}{2} \Omega_m \mu(k,a) \delta = 0.
\end{eqnarray}
Defining the linear growth function
\begin{equation}
    f(t) \equiv \frac{d~\text{ln} ~D_+(t)}{d~\text{ln} ~a(t)}, 
\end{equation}
the MG growth equation is:  
\begin{eqnarray} \label{f_MG}
    f^{'} + f^{2} + \left[ 2 - \frac{3}{2}  \left ( 1 + w_\phi (1- \Omega_m - \Omega_\Lambda) + w_{\Lambda} \Omega_{\Lambda} \right)  \right] f - \frac{3}{2} \Omega_m \mu(k,a)  = 0.
\end{eqnarray}
eq. (\ref{delta_MG}), or eq. (\ref{f_MG}) are valid in general, with $w_{\Lambda}$ provoking a friction term that comes from the  background expansion that in general can be different from $\Lambda$CDM and $\mu(k,a)$ leads to an enhancement of the effective matter density's contrast. Both terms change the structures' collapse with respect to $\Lambda$CDM. Given that the source of the Poisson equation for MG depends on scale through the function $\mu(k,a)$, the growth rate and growth functions inherit this property too. However, the scale dependence complicates its comparison with the  measured growth function or growth rate. One approach is to compute it using $k$-bins.
\begin{figure}
 	\begin{center}
    \includegraphics[width=2.6 in]{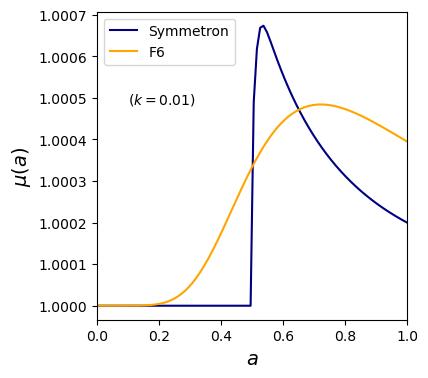}
    \includegraphics[width=2.6 in]{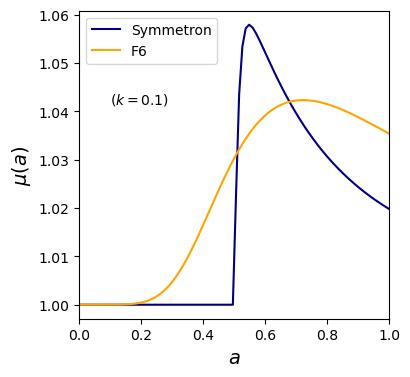}
 	\caption{Function $\mu(k,a)$ for Hu-Sawicki and Symmetron models. Left panel evaluated at $k=0.01$ and the right panel is evaluated at $k=0.1$, in units of $h~{\rm Mpc^{-1} }$. The curves correspond to Symmetron (blue) and F6 (orange) models. The transition value for Symmetron is $z_{ssb}=1$}
  \label{fig:mu}
 	\end{center}
\end{figure}
In the Symmetron model one can choose the redshift transition value that we chose $z_{ssb}=1$ as in \cite{Hinterbichler:2011ca, Brax_2012}, and for $z>z_{ssb}$ the evolution follows the $\Lambda$CDM model.  In Fig. \ref{fig:mu} we show the $\mu(k,a)$ function corresponding to the $f(R)$ and Symmetron models for different values of $k$ in units of $h~{\rm Mpc^{-1}}$. We can directly see that the impact of the function in the Symmetron model is present at different scales and the behavior remains the same over time.

It is interesting to note that $\mu(k,a)$ corresponding to the Symmetron model is greater than F6 model for the two $k$ values at the  transition value, and just after this point the Symmetron function reaches its maximum value but quickly decreases with $a$. And, this behavior will be therefore observed in the linear growth function. Another important thing, is that the deviation of the Symmetron model with respect to $\Lambda$CDM is of the order of F6, which means the model is viable, since the effects for F4 and F5 models are stronger, but they are practically ruled out by observations \cite{He_2018}. Thus, the relevance of Fig. \ref{fig:mu} is that it shows clearly that the $\mu(k,a)$ function represents an important contribution in the differential equation for $f(k,z)$ and it will be relevant in the fk-kernel approximation \cite{Aviles:2021que}, as can be seen in the implemented codes: fkpt \cite{Rodriguez_Meza_2024} for HS-$f(R)$ and FOLPS-nu\footnote{The MG models in this work were implemented in FOLPS-nu.} \cite{Noriega:2022nhf} for neutrinos. 

The fk-kernel approximation consists of taking the $\Lambda$CDM kernels, but keeping the scale dependence in the linear growth rate. As the authors show in \cite{Noriega:2022nhf} this approximation works for the $\Lambda$CDM plus neutrinos and also works for HS-$f(R)$ as is shown in \cite{Rodriguez_Meza_2024}. In addition, this approximation is useful to reduce the computational cost of the loop integral calculations of the nonlinear contributions.

In Fig. \ref{fig:f_z} we show $f(k,z)$ as function of $z$ for two different $k$-bins: ($0.01 \leq k \leq 0.05$) and ($0.06 \leq k \leq 0.1$) in units of $h~\text{Mpc}^{-1}$, c.f. similar results in Fig. 2 of ref. \cite{Mirzatuny:2019dux}, where the authors show analogous plots by using a fitting formula instead of solving the differential equation that accounts for the exact contribution of the scale dependence of the function $\mu(k,a)$. 
 \begin{figure}
 	\begin{center}
 	\includegraphics[width=2.2 in]{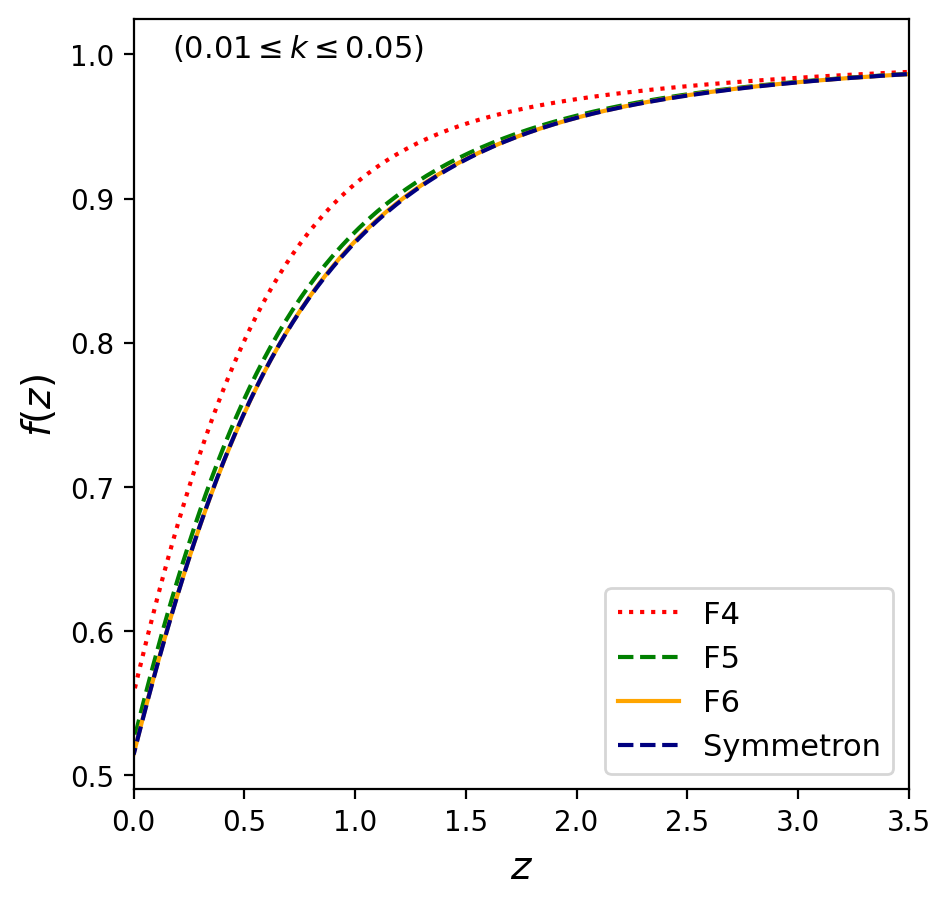}
 	\includegraphics[width=2.2 in]{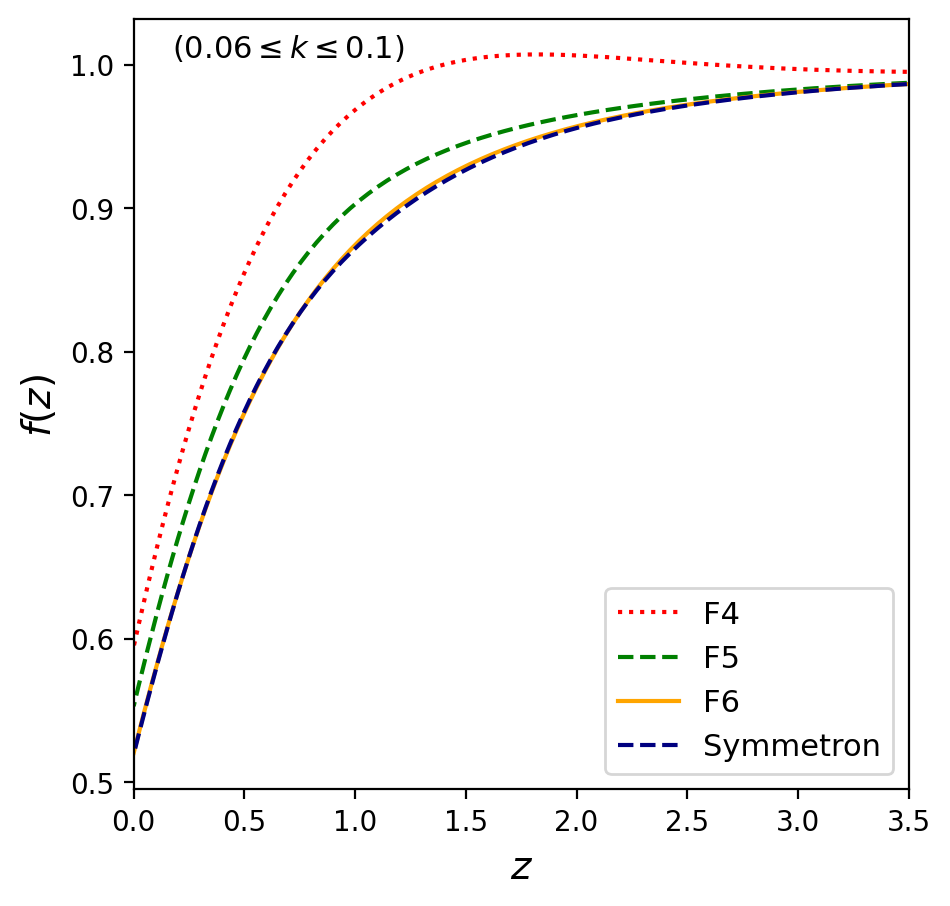}
    \includegraphics[width=2.2 in]{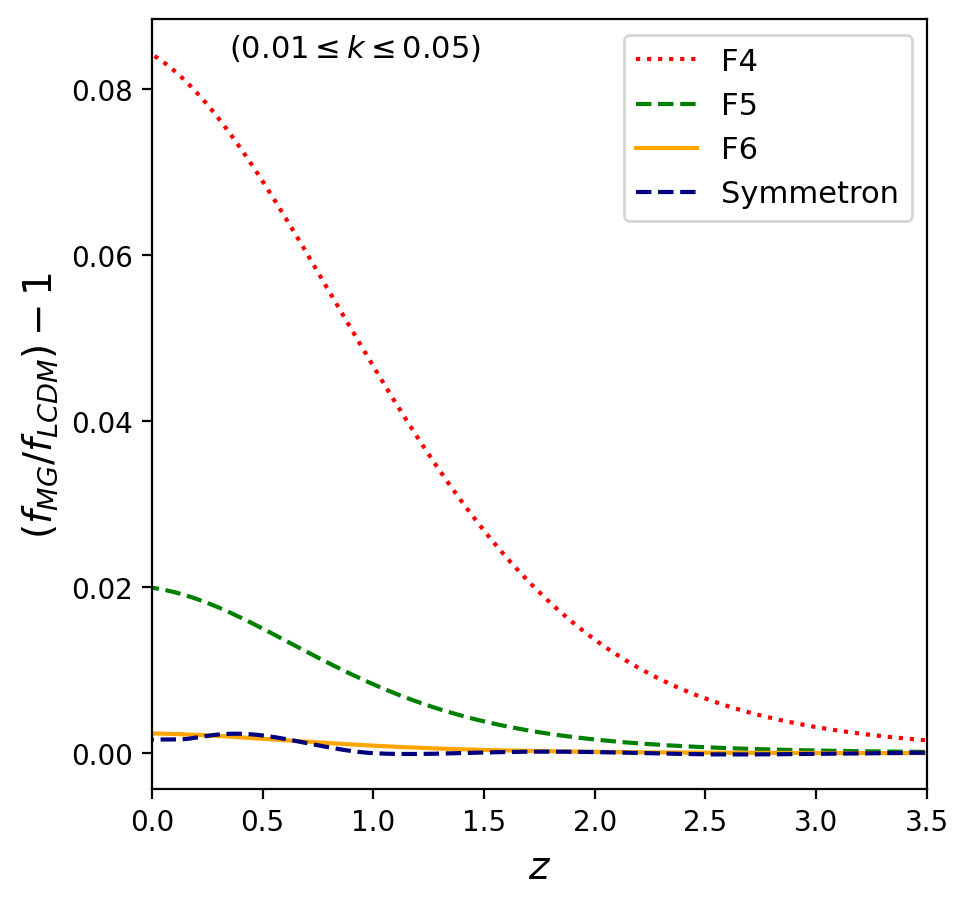}
    \includegraphics[width=2.2 in]{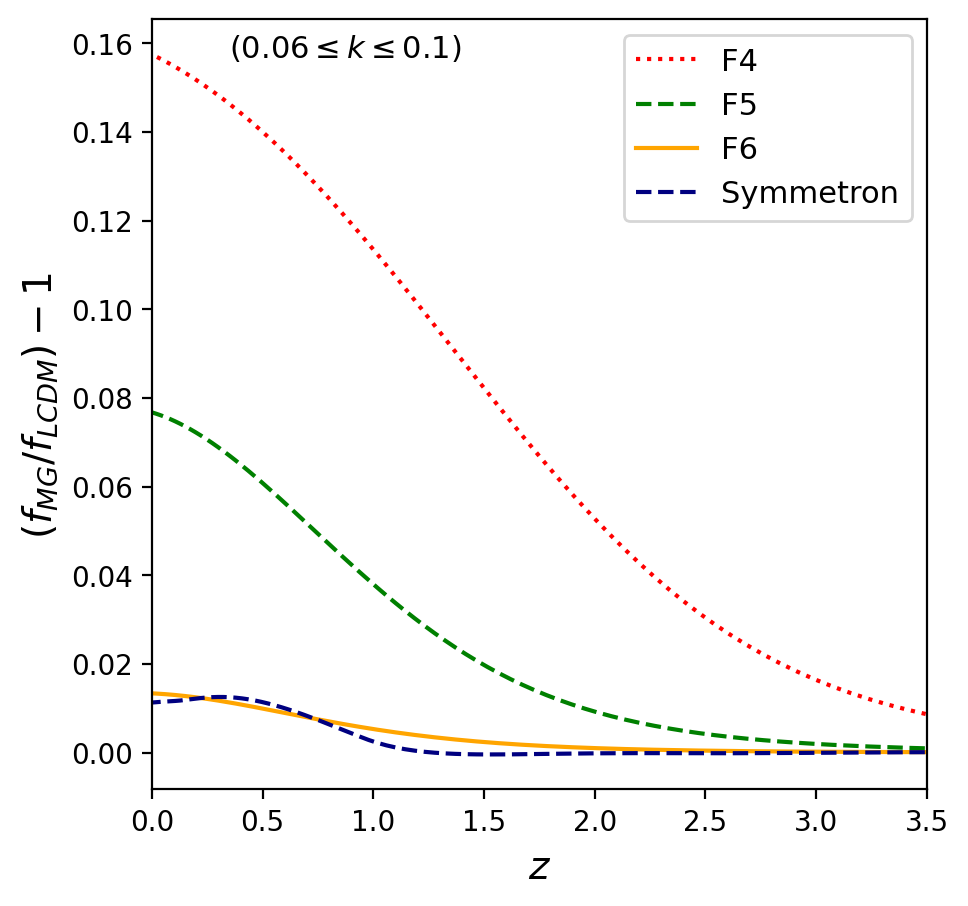}
 	\caption{$f(k,z)$ for Hu-Sawicki $f(R)$ and Symmetron models for two $k$-bins.}
  \label{fig:f_z}
 	\end{center}
 \end{figure}
In our case, we show the $f(k,z)$ for Hu-Sawicki, Symmetron, and $\Lambda$CDM models. Top panels show $f(k,z)$ functions and bottom panels show the relative difference with respect to $\Lambda$CDM.

All modified gravity models evaluated at the value of  $k=10^{-3} h/{\rm Mpc}$ coincide with $\Lambda$CDM model, so we did not plot these results, i.e., we found that for F4 the maximum deviation is around $0.01 \%$ at $z=0$, while for the rest of models found a maximum deviation smaller than $0.002 \%$. On the other hand, for the case of the first $k$-bin considered, the F4 model begins to depart from the reference curve ($\Lambda$CDM) with a deviation around $8 \%$ at $z=0$ and around $5 \%$ at $z=1$. Though  the MG footprint is small, it can potentially be detected with data from DESI survey. The rest of the models have a deviation less than $2 \%$ for $z \geq 0$ and its effect is challenging to be detected.

For the second $k$-bin considered, $0.06 \leq k \leq 0.1$, we can see on bottom panel for F4 the deviation extends to nearly $16 \%$ at $z=0$ and is around $12 \%$ at $z=1$; for the F5 model, the deviation is around $8 \%$ at $z=0$ and is lesser than $4 \%$ for $z>1$ ; for Symmetron and F6 models the deviation is smaller than $2 \%$ for $z \geq 0$. The deviation of the Symmetron and F6 models are always of the same order in both $k$-bins considered. Also, this is consistent with one result in \cite{Ruan:2021wup} where the authors showed that the behavior of the matter PS for Symmetron is similar to $f(R)$ at $z=0$. We can see from bottom panels of Fig. \ref{fig:f_z} that whereas the behaviors of $f(z)$ function for $f(R)$ model continue increasing all the way to smaller redshift, the Symmetron first increases until it reaches a maximum and then decreases with respect to $\Lambda$CDM that is consistent with the behavior shown in Fig. \ref{fig:mu} for $\mu(k,a)$. 

The Symmetron model shows its transition at $z=z_{ssb} $, this is when the MG begins to act and at this point a steep increase in the growth rate is generated. The model shows its maximum deviation around $z=0.5$ with some more growth than the F6 model at this redshift, decreasing a little less than F6 at $z=0$; these difference are less than $1 \%$. Such behavior occurs for all $k$-values and it can be seen better on the bottom panels of Fig. \ref{fig:f_z}, where the difference with respect to $\Lambda$CDM was plotted and it is clear that the maximum deviation occurs around $z=0.5$. 
Conversely, in Fig. \ref{fig:f_k} we show the linear growth rate as a function of $k$ for three $z$-values ($0.8, 0.5, 0$), c.f. similar results in Fig. 3 of Ref. \cite{Mirzatuny:2019dux} for the $f(R)$ model. 
\begin{figure}
 	\begin{center}
  \includegraphics[width=1.9 in]{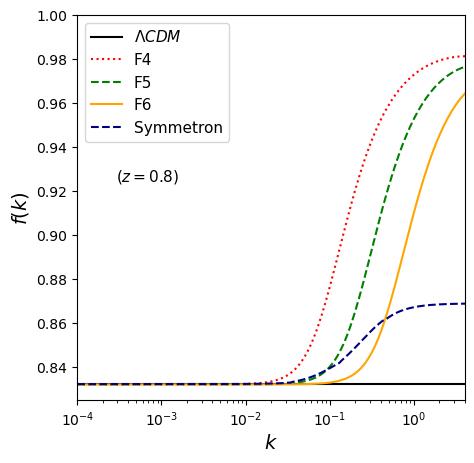}
 	\includegraphics[width=1.9 in]{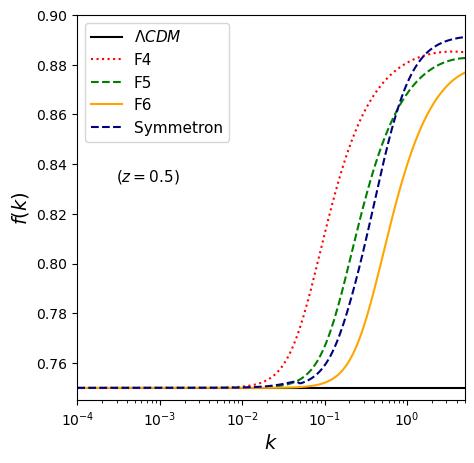}
  \includegraphics[width=1.9 in]{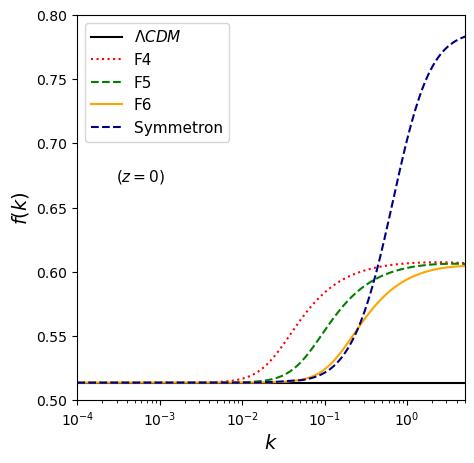}
 	\caption{Comparison of the  growth function for $f(R)$ and Symmetron models as function of $k$ in units of $h~{\rm Mpc^{-1} }$, for the $z$-values ($ 0.8, 0.5, 0$). The black line is the value for $\Lambda$CDM.}
  \label{fig:f_k}
 	\end{center}
\end{figure}
In our case, we can see the contribution of the $\mu(k)$ for each $z$-value, showing the relevance of $k$-dependence in the growth rate. The previous comments are considering the range $k \in [10^{-4},0.1]$, since $k>0.1$ is the nonlinear regime and the theory does not work. The broader interval shown in three cases is just for visualization. We can see the same behavior that in Fig. \ref{fig:f_z}, that is at $z=0$ the deviation is of the order of F6 for the $k$-values of interest, being greatest for $z$-values near to $z=z_{ssb}$ and recovering $\Lambda$CDM (Einstein-de Sitter phase) for $z>z_{ssb}$.

The growth rate in the vanilla $\Lambda$CDM model is well approximated by
\begin{equation}
    f(a) = \Omega_m^\gamma(a),
\end{equation}
with the growth index $\gamma=0.55$. Even when studying models beyond $\Lambda$CDM this formula is valid, but taking different constant values for $\gamma$ or parametrizations of it \cite{Linder_2005,Linder_2007, Bean:2010zq, Ferreira_2010} also for models with different gravitational theories \cite{Silvestri_2013}. Recently, it has been shown that the usual value of $\gamma$ is not correct for the dominated era of dark energy \cite{Nguyen_2023} and with upcoming surveys it will be possible to test the growth rate and put constraints on different MG models \cite{SPT:2024adw} or using RSD measurements \cite{Chapman_2023}.
As discussed in this Section, the growth rate for a modified gravity model is obtained by solving the differential equation (\ref{delta_MG}) or (\ref{f_MG}), resulting in a scale-dependent function not only dependent on time. With this in mind and similarly to Ref. \cite{Mirzatuny:2019dux}, we can propose a parametrized formula given by:
\begin{equation}
    f(k) = c_1 \text{Tanh}^2(c_2 k) + c_3,
    \label{fk_parameterization}
\end{equation}
where $c_i$ are constants and can be fixed to emulate F6 or Symmetron model. In particular the constant $c_3$ plays the role of $f_0$ for the MG model in consideration and is exactly the growth rate at $k=0$, as in the $\Lambda$CDM model, $ c_3 =f_0 = \Omega^{0.55}_m (k=0)$. The constants $c_{1}$ and $c_{2}$ depend mostly on $\mu(k,a)$, but we cannot determine an analytical dependence  because the part of the Tanh$^2$ function that emulates $f(k,t)$ is somewhat degenerate in the parameters $c_{1}$ and $c_{2}$. From Eq. (\ref{f_MG}), the scale dependence is given by the function $\mu(k,a)$, which characterizes the gravity model. And in both models, HS-$f(R)$ and Symmetron, the scale dependence is the same, as seen in equations (\ref{mu_fr}) and (\ref{mu_symmetron}), respectively.

On the left panel of Fig. \ref{fig:fk_parameterization} we plot $f(k)$ at $z=0$ for Symmetron and the parametrization formula, on the middle panel we plot $f(k)$ at $z=0$ for F6 and the parametrization formula, and on the right panel we plot the deviation of the parametrization with respect to each model. The values used for Symmetron were $\{ c_1,c_2,c_3\}= \{ 0.265,3.55,0.5048\}$ and for F6: $\{ c_1,c_2,c_3\}= \{ 0.265,3.05,0.50485\}$. We solved the differential equation for $f$ in the range $k \in [10^{-4},10^3]$. The parametrization have a maximum deviation lesser than $0.6\%$  with respect to the MG models up to $k=0.2$ $h$/Mpc. Thus, eq. (\ref{fk_parameterization}) could be a good approximation to model the scale dependence of the linear growth function in both models. Furthermore, the simplicity of the parametrization can be useful for the calculations of nonlinear contributions because solving differential equation (\ref{delta_MG}) or (\ref{f_MG}) can be avoided, when inferring cosmological parameters in an MCMC computation.
\begin{figure}
 	\begin{center}
 	\includegraphics[width=1.9 in]{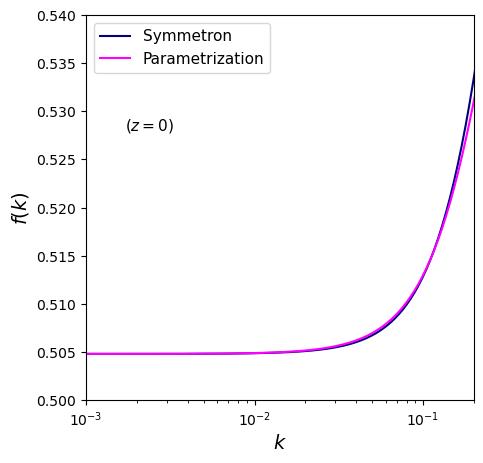}
    \includegraphics[width=1.9 in]{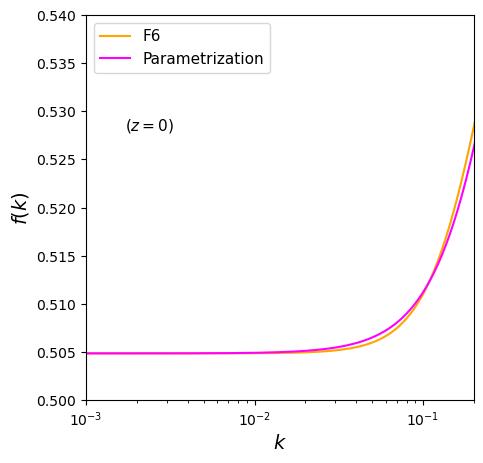}
    \includegraphics[width=1.9 in]{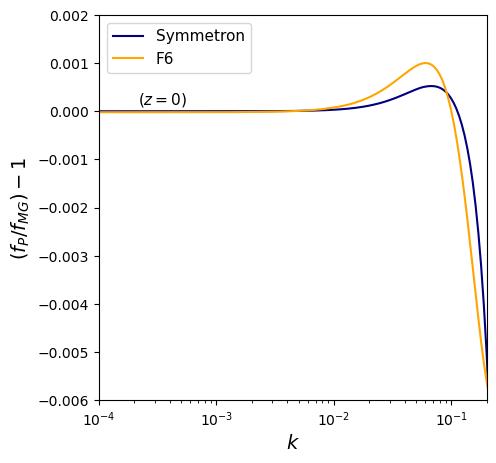}
 	\caption{Comparison of the linear growth rate for HS-$f(R)$, Symmetron, and their parametrization (eq. \ref{fk_parameterization}) as function of $k \in [10^{-3},0.2]$ $h~{\rm Mpc^{-1} }$ at $z=0$. In the right panel we show the relative difference for each model, where the f$_{\text{MG}}$ label is for the modified gravity model and f$_{p}$ is for the parametrization. } 
 	\label{fig:fk_parameterization}
 	\end{center}
\end{figure}
\end{section}
\begin{section}{Nonlinear MG contributions} \label{sect:nonlinear_MG}
The MG theories studied are determined by the modified Poisson equation \cite{Koyama:2009me,Aviles:2018saf},
\begin{equation}
    \frac{1}{a^2} \nabla^2 \Phi = 4 \pi G \bar{\rho} \delta - \frac{1}{2a^2} \nabla^2 \varphi,
    \label{PoissonMG}
\end{equation}
where the bar above $\rho$ stands for the dark matter energy density of the background, $ \delta$ its perturbation, and $\Phi$ the gravitational potential, cf. eq. (\ref{metric}). The scalar field ($\varphi$) is coupled to the dark matter energy density through the Klein-Gordon equation, and when the quasi-static approximation is taken and considering the case of small-scale mode interactions, it can be written as \cite{Aviles:2017aor, Koyama:2009me}
\begin{equation}
    (3+2w_{\text{BD}}) \frac{1}{a^2} \nabla^2 \varphi = -8 \pi G \bar{\rho} + \text{NL},
    \label{scalarfieldeq}
\end{equation}
with NL being the possible non-linearities. Introducing eq. (\ref{scalarfieldeq}) into the Poisson equation, and expanded in Fourier space one obtains \cite{Aviles:2020wme}, 
\begin{equation}
    - \frac{k^2}{a^2} \Phi = A(k,t) \delta(\Vec{k}) + S(\Vec{k}),
    \label{poisson}
\end{equation}
where the function $A(k,t)$ determines the linear model, whereas $S(\Vec{k})$ represents the nonlinear contribution. One identifies the function $A(k,t)$ in the usual way by the function $ A(k,t)= 4\pi G \overline{\rho} \mu (k,t)$, where the function $\mu (k,t)$ is more widely employed to parameterize the linear effects of different MG models. The origin of the $S(\Vec{k})$ function in the case of MG models is from non-linearities of the Klein-Gordon equation and it is responsible for the screening mechanism that enables to recover General Relativity at small scales. One can express $S(\Vec{k})$ in a Taylor series in Fourier space as shown in \cite{Aviles:2020wme},
\begin{equation}
    S(\Vec{k}) = \frac{1}{2} \int\limits_{\Vec{k}_{12}=\Vec{k}} \mathcal{S}^{(2)} (\Vec{k}_{1}, \Vec{k}_{2}) \delta(\Vec{k}_{1}) \delta(\Vec{k}_{2}) 
    +
    \frac{1}{6} \int\limits_{\Vec{k}_{123}=\Vec{k}} \mathcal{S}^{(3)} (\Vec{k}_{1},\Vec{k}_{2},\Vec{k}_{3}) \delta(\Vec{k}_{1}) \delta(\Vec{k}_{2}) \delta(\Vec{k}_{3}) + \cdots,
    \label{S}
\end{equation}
with each $\mathcal{S}^{(n)}$ term being determined once we specify the theory. 

Inserting eq. (\ref{poisson}) into the Euler and continuity equations, these are, in Fourier space given by: 
 \begin{align}
 \frac{1}{H}  \frac{\partial\delta(\vk)}{\partial t} - f_0 \theta(\vk) &=  f_0  \ikk \alpha(\vk_1,\vk_2) \theta(\vk_1) \delta(\vk_2), \label{FScontEq}\\
  \frac{1}{H}  \frac{\partial f_0 \theta(\vk)}{\partial t} + \left( 2+ \frac{\dot{H}}{H^2}\right) f_0 \theta(\vk)
 &- \frac{A(k)}{H^{2}} \delta(\vk)  -\frac{S(\vk)}{H^2} =   f_0^2 \ikk \beta(\vk_1,\vk_2) \theta(\vk_1) \theta(\vk_2), \label{FSEulerEq}
 \end{align}
 where the  $\theta(\vk)$ function is the velocity field and $\delta(\vk)$ is the matter field. Identifying the functions
 \begin{equation}
  \alpha(\vk_1,\vk_2) = 1+\frac{\vk_{1}\cdot\vk_2}{k_1^2},  \qquad  \beta(\vk_1,\vk_2) = \frac{k_{12}^2(\vk_1\cdot\vk_2)}{2 k_1^2 k_2^2}. 
 \end{equation}
Given the analytical expressions for $A(k,t)$ and $S(\Vec{k})$ we can find perturbative solutions of $n$-orders using eqs. (\ref{S}-\ref{FSEulerEq}). 
\subsection{Hu-Sawicky f(R) }

For Hu-Sawicky $f(R)$ model the $A(k,t)$ and $S(\vk)$ functions are well known. Then, from the calculation of eq. (\ref{S}) we can rewrite $S(\Vec{k})$ as follow,
\begin{align}
S(\Vec{k}) &= -\frac{1}{6}  \frac{k^2}{k^2 +  a^2m^2(a)}  {\delta \cal I},
\label{S_Eq_fR}
\end{align}
where $m\equiv \sqrt{M_1/3}$, $M_1(k)$ function encodes the gravitational pull at linear level and can be absorbed by the function $A(k,t)$  and ${\delta \cal I}$ only contains $M_{2,3}$ functions responsible for the screening behavior and is given by:
\begin{align}
\delta \mathcal{I}(\delta f_{R})
&= \frac{1}{2} \int \frac{d^3 k_1 d^3 k_2}
{(2 \pi)^3} \delta_D(\vk -\vk_{12}) M_2(\vk_1, \vk_2)
\delta f_{R}(\vk_1) \delta f_{R}(\vk_2) \nonumber\\
&\quad + \frac{1}{6}
\int \frac{d^3 k_1 d^3 k_2 d^3 k_3}{(2 \pi)^6}
\delta_D(\vk - \vk_{123}) M_3(\vk_1, \vk_2, \vk_3)
\delta f_{R}(\vk_1) \delta f_{R}(\vk_2) \delta f_{R}(\vk_3) + \cdots.
\label{delta_I}
\end{align}
The $M_{2,3}$ functions are second and third order functions that are responsible for the screening mechanism and for this modified gravity model are given by:
\begin{eqnarray}
    M_2(a) &=& \frac{9}{4} \frac{H_0^2}{|f_{R0}|^2} \frac{(\Omega_ma^{-3} + 4\Omega_\Lambda )^5}{(\Omega_m + 4\Omega_\Lambda)^4}, 
    \label{M2} \\
    M_3(a) &=& \frac{45}{8} \frac{H_0^2}{|f_{R0}|^3} \frac{(\Omega_ma^{-3} + 4\Omega_\Lambda )^7}{(\Omega_m + 4\Omega_\Lambda)^6}.
    \label{M3}
\end{eqnarray}
We note that these functions are only background quantities dependent, but this is a special feature of the Hu-Sawicki model.

\subsection{Symmetron}
In the Symmetron case we can obtain $A(k,t)$ from the $\mu(k,t)$ given in equation (\ref{mu_symmetron}).
\begin{equation}
    A(k,t) = 4 \pi G \bar{\rho} \left( 1 + \frac{2 \beta^2(a)}{A(\phi)} \frac{ k^2 }{ k^2 + a^2 m^2(a) } \right).
\end{equation}
It is important to mention that Symmetron model also features a screening and anti-screening mechanism \cite{Aviles:2018qot}. Similarly as in the $f(R)$ model we can compute the $M_{2,3}$ functions responsible for the screenings, but unlike $f(R)$ in the Symmetron model these functions depend on both scale and time (for more details on the following functions, see \cite{Aviles:2018qot}). We will omit the time argument for simplicity visualization in the equations:

\begin{eqnarray}
    M_2(\vk_1,\vk_2) &=& \frac{2 A \beta_2 A_0 }{\beta^2} \mathcal{K}_{\chi^\delta}^{(1)}(\vk_1) - \frac{ A^2 \kappa_3}{4 \beta^3},
    \label{M2_symmetron} \\
    M_3(\vk_1, \vk_2, \vk_3) &=& \frac{3 A \beta_2 A_0 }{\beta^2} \mathcal{K}_{\chi^\delta}^{(2)}(\vk_1, \vk_2) - \frac{ 3 A^2 \beta_3 A_0}{2 \beta^3}\mathcal{K}_{\chi^\delta}^{(1)}(\vk_1) + \frac{ A^3 \kappa_4}{8 \beta^4},
    \label{M3_symmetron}
\end{eqnarray}
where $\mathcal{K}_{\chi^\delta}^{(1,2)}$ are kernels given by:
\begin{eqnarray}
    \mathcal{K}_{\chi^\delta}^{(1)} (\vk_1) &=& \frac{A}{6a^2\beta^2} (k_1^2 +m^2 a^2),
    \label{kappa1}  \\
    \mathcal{K}_{\chi^\delta}^{(2)} (\vk_1, \vk_2) &=& \frac{1}{2A_0} M_2^{\text{FL}}(\vk_1, \vk_2).
    \label{kappa2}
\end{eqnarray}
The frame-lagging function $M_2^{\text{FL}}$ is
\begin{equation}
    M_2^{\text{FL}} = M_2(\vk_1,\vk_2) + \frac{9 A}{4 \beta^2 A_0^2} \mathcal{K}_{\text{FL}} (\vk_1,\vk_2) \Pi(\vk_1) \Pi(\vk_2),
    \label{M2FL}
\end{equation}
$\mathcal{K}_{\text{FL}}$ is the frame-lagging kernel of second order that depends on $A(k,t)$ function; $\Pi(k_i) \equiv \frac{1}{3a^2} (3k_i^2 + M_1a^2)$; and functions $\beta_n, \kappa_n$ defined in \cite{Brax_2013}: 
\begin{align*}
    & \beta_n = M^n_{pl} ~ \frac{d^n A(\phi)}{d\phi}, \\
    & \kappa_n = M^{n-2}_{pl} \left( \frac{d^nV(\phi)}{d\phi} + \bar{\rho} ~ \frac{d^nA(\phi)}{d\phi} \right). 
\end{align*}
The $M_{2,3}$ functions encode the screening and anti-screening mechanisms \cite{Aviles:2018qot} and they are necessary to compute numerically the growth functions for the construction of the power spectrum up to 1-loop. The frame-lagging term arises when the transformation of the Klein-Gordon equation from Standard Eulerian Perturbation Theory (SPT) to Lagrangian Perturbation Theory (LPT) is performed and is necessary to recover $\Lambda$CDM on large scales \cite{Aviles:2017aor}.

\subsection{Full one-loop power spectrum } \label{sect:full 1-loop PS}
In the study of the PT when one considers a general MG theory, it is necessary to take into account the scale dependence in the growth rate even in the linear regime, as we showed in Section \ref{sect:scale_dep_growth}. 
In general, the PS for PT up to 1-loop is given by \cite{Aviles:2020wme}:

\begin{equation}
    P^{1-loop}_{ab}(k) = P^L_{ab}(k) + P^{22}_{ab} (k) + P^{13}_{ab} (k),
\end{equation}
where $a,b$ refer to matter or velocity fields and $P^L_{ab}(k)$ is the linear power spectrum
\begin{equation} \label{Plinear}
 P^L_{\delta\delta}(k) \equiv P_L(k) , \quad
 P^L_{\delta\theta}(k) =  \frac{f(k)}{f_0} P_L(k), \quad
 P^L_{\theta\theta}(k) =  \left(\frac{f(k)}{f_0} \right)^2 P_L(k),
 \end{equation}
whit $f_0=f(k=0,t)$ and $P^{22}_{ab} (k)$ and $P^{13}_{ab}$ are the pure 1-loop contributions \cite{Aviles:2020wme,Aviles:2018saf}

\begin{align}
P_{\delta\delta}^{22}(k) &=   2 \ip \big[F_2(\vp,\vk-\vp)\big]^2 P_L(p) P_L(|\vk -\vp|), \\
P_{\delta\theta}^{22}(k) &=   2 \ip F_2(\vp,\vk-\vp)G_2(\vp,\vk-\vp) P_L(p) P_L(|\vk -\vp|), \\
P_{\theta\theta}^{22}(k) &=   2 \ip \big[G_2(\vp,\vk-\vp)\big]^2 P_L(p) P_L(|\vk -\vp|), \\
P_{\delta\delta}^{13}(k) &=   6  P_L(k) \ip F_3(\vk,-\vp, \vp) P_L(p),\\
P_{\delta\theta}^{13}(k) &=   3  P_L(k) \ip \big[F_3(\vk,-\vp, \vp)G_1(\vk) + G_3(\vk,-\vp, \vp) \big] P_L(p),\\
P_{\theta\theta}^{13}(k) &=   6  P_L(k) \ip G_3(\vk,-\vp, \vp)G_1(\vk) P_L(p). 
\end{align}
There are two approaches to obtain the SPT $F_n$ and $G_n$ kernels, the first is to calculate them using Euler and continuity equations and find $n$-order solutions by solving iteratively. The second way is to compute the LPT kernels, then doing the translation into the SPT kernels. Actually, the later approach was developed in \cite{Aviles:2018saf,Matsubara:2011ck} where they do the map from LPT to SPT kernels. In fact, the use of LPT is more convenient in the study of baryon acoustic oscillations (BAO) features and the correlation function around the acoustic peak. Furthermore, unlike SPT, the PS of LPT can be transformed to Fourier space to obtain the correlation function. However, SPT follows better  the broadband form of the PS and, in the case of LPT, it is observed that at small scales the PS decays quickly causing a deficit in the formation of structures at these scales. Both approaches have pros and cons.  

Up to second order the $F_2$ and $G_2$ kernels are given by \cite{Aviles:2021que,Aviles:2020wme,Rodriguez_Meza_2024}:
\begin{align}
F_2(\vk_1,\vk_2) &= \frac{1}{2} + \frac{3}{14}\A + \left( \frac{1}{2} - \frac{3}{14}\B  \right)   \frac{(\vk_1\cdot\vk_2)^2}{k_1^2 k_2^2}
        + \frac{\vk_1\cdot\vk_2}{2 k_1k_2} \left(\frac{k_2}{k_1} + \frac{k_1}{k_2} \right), \label{F2_kernel}\\
G_2(\vk_1,\vk_2) &= \frac{3\A(f_1+f_2) + 3 \dot{\A}/H }{14 f_0} +
\left(\frac{f_1+f_2}{2 f_0} - \frac{3\B(f_1+f_2) + 3 \dot{\B}/H }{14 f_0}\right) \frac{(\vk_1\cdot\vk_2)^2}{k_1^2 k_2^2} \nonumber\\
&\quad + \frac{\vk_1\cdot\vk_2}{2 k_1k_2} \left( \frac{f_2}{f_0}\frac{k_2}{k_1} + \frac{f_1}{f_0}\frac{k_1}{k_2} \right), \label{G2_kernel}
\end{align}
where $f_{1,2} = f(\vk_{1,2})$. We can see the presence of $\mathcal{A}$ and $\mathcal{B}$ functions that are time and scale dependent
\begin{equation} \label{AandBdef}
 \mA(\vk_1,\vk_2,t) = \frac{7 D^{(2)}_{\mA}(\vk_1,\vk_2,t)}{3 D_{+}(k_1,t)D_{+}(k_2,t)}, 
 \qquad \mB(\vk_1,\vk_2,t) = \frac{7 D^{(2)}_{\mB}(\vk_1,\vk_2,t)}{3 D_{+}(k_1,t)D_{+}(k_2,t)},
\end{equation}
these functions arise because the relation $f^2=\Omega_m(t)$ does not hold, with $D^{(2)}_{\mA}(\vk_1,\vk_2,t)$ and $D^{(2)}_{\mB}(\vk_1,\vk_2,t)$ the second order growth functions being solution of the linear second order differential equations \cite{Aviles:2017aor}:
\begin{align}
\big(\T - A(k)\big)D^{(2)}_{\mA} &= \Bigg[A(k) + (A(k)-A(k_1))\frac{\vk_1\cdot\vk_2}{k_2^2} + (A(k)-A(k_2))\frac{\vk_1\cdot\vk_2}{k_1^2} \nonumber\\
             &     \qquad   +  \mathcal{S}^{(2)}(\vk_1,\vk_2) \Bigg]  D_{+}(k_1)D_{+}(k_2), \label{DAeveq} \\
\big(\T - A(k)\big)D^{(2)}_{\mB} &= \Big[A(k_1) + A(k_2) - A(k) \Big]  D_{+}(k_1)D_{+}(k_2). \label{DBeveq}
\end{align}
For third order kernels the computation is more cumbersome in SPT and it is more convenient to do the map from the known LPT kernels, they are given by \cite{Aviles:2021que,Aviles:2020wme,Rodriguez_Meza_2024}:
\begin{align}
F_3(\vk,-\vp,\vp) &= \frac{1}{6} \Gamma_3(\vk,-\vp,\vp) -\frac{1}{6} \frac{(\vk\cdot\vp)^2}{p^4}  
 +  \frac{1}{7}\frac{(\vk \cdot (\vk-\vp)) (\vk\cdot\vp)}{|\vk-\vp|^2 p^2}\Gamma_2(\vk,-\vp), \label{F3kernel}\\
 G_3(\vk,-\vp,\vp) &=  
   \frac{1}{2} \Gamma^{f}_3(\vk,-\vp,\vp)   +  \frac{2}{7} \frac{\vk \cdot \vp}{p^2} \Gamma^{f}_2(\vk,-\vp)   + \frac{1}{7}  \frac{f(p)}{f_0}  \Gamma_2(\vk,-\vp) \frac{\vk \cdot (\vk-\vp)}{|\vk-\vp|^2} \nonumber\\
&\quad    -  \frac{1}{7}   \left[ 2 \Gamma_2^f(\vk,-\vp) + \frac{f(p)}{f_0} \Gamma_2(\vk,-\vp) \right]\left[1 -\frac{(\vp \cdot (\vk-\vp))^2}{p^2|\vk-\vp|^2}  \right]\nonumber\\
&\quad - \frac{1}{6} \frac{(\vk \cdot \vp)^2}{p^4} \frac{f(k)}{f_0},\label{G3kernel}
\end{align}
where
\begin{equation}\label{LDDfkernels}
 \Gamma^f_n(\vk_1,\cdots,\vk_n,t) \equiv  \Gamma_n(\vk_1,\cdots,\vk_n,t) \frac{f(k_1)+\cdots+f(k_n)}{n f_0} + \frac{1}{n f_0 H} \dot{\Gamma}_n(\vk_1,\cdots,\vk_n,t).
\end{equation}
The $F_n$ and $G_n$ PT kernels are the way to compute the exact 1-loop contributions. However, the use of them increases the computational cost for the scale-dependent MG case, which is amplified when we estimate cosmological parameters using an MCMC code since differential equations have to be solved to find them.

On the other hand, to compute the galaxy PS from the dark matter PS, one introduces redshift space distortions (RSD), that account for the bias and the galaxy peculiar velocity. This behavior can be described by a velocity moments generating function \cite{Vlah:2018ygt, Scoccimarro:2004tg} that can be expanded in  many different ways to yield different approaches to RSD modeling \cite{Vlah:2018ygt}. We use this approach and further expand the perturbative approach with the use of the functions $I_n^m(k)$ \cite{Aviles:2020wme}, to obtain  
\begin{eqnarray}
    P_s^{\text{ME}}(k,\mu) = \sum_{m=0}^\infty \sum_{n=0}^m \mu^{2n} f_0^m I_n^m(k).
    \label{PME}
\end{eqnarray}
Furthermore, when we take account the EFT corrections and the shot-noise terms, we can write the full 1-loop expression for the PS, that is, 
\begin{equation}
    P_s(k,\mu) = P_s^{\text{ME}}(k,\mu) + P_s^{\text{EFT}}(k,\mu) + P_s^{\text{shot}}(k,\mu),
    \label{Ps}
\end{equation}
where the first term can be obtained by expanding eq. (\ref{PME}) with $n=2$, this results in
\begin{equation}
    P_s^{\text{ME}}(k,\mu) = P_{\delta \delta}(k) + 2f_0 \mu^2 P_{\delta \theta}(k) + f_0^2 \mu^4 P_{\theta \theta}(k) + A^{\text{TNS}} (k,\mu) + D(k,\mu).
    \label{PsME}
\end{equation}
The function $A^{\text{TNS}}$ corresponding to the corrections to the non-linear Kaiser model \cite{Taruya:2010mx} is
\begin{align}
A(k,\mu)= 2 \mu k f_0 \ip  \frac{\vp\cdot \vhn}{p^2} B_\sigma(\vp,-\vk,\vk-\vp) \, , 
\end{align}
where $B_\sigma$ is the bispectra contributions given in \cite{Aviles:2020wme}. The rest of the velocity moments contributions are placed in $D$ function
\begin{align} \label{D_k_mu}
 D(k,\mu) = B(k,\mu) + C(k,\mu)  -(\mu k f_0 \sigma_v)^2 P^K_s(k,\mu).
\end{align}
wherein \cite{Taruya:2010mx}
\begin{align}
 B(k,\mu) &= (k\mu f_0)^2  \ip F(\vp)F(\vk-\vp) , \\
  F(\vp) &= \frac{\vp\cdot\vhn}{p^2}\Bigg[ P_{\delta\theta}(p) + f_0 \frac{(\vp\cdot\vhn)^2}{p^2} P_{\theta\theta}(p) \Bigg], 
\end{align}
and 
\begin{equation}
 C(k,\mu) = (k\mu f_0)^2  \ip  \frac{(\vp\cdot\vhn)^2}{p^4}P_{\theta\theta}(p) P^K_s(|\vk-\vp|, \mu_{\vk-\vp}).
\end{equation}
The second and third terms of eq. (\ref{Ps}) are given by,
\begin{align}
    & P_s^\text{EFT} (k, \mu) = \big(\alpha_0 + \alpha_2 \mu^2 + \alpha_4 \mu^4 + \alpha_6 \mu^6\big) k^2 P_L(k) + \tilde{c} \big(\mu k f_0 \big)^4   P^K_s (k,\mu), \label{PEFT} \\
    &     P_s^\text{shot}(k, \mu) = \frac{1}{\bar{n}} \big( \alpha_0^\text{shot} + k^2 \mu^2 \alpha^\text{shot}_2 \big),
    \label{Pshot}
\end{align}
with $\bar{n}$ the average number density of galaxies. For the case of a Poisson distribution, $\alpha_0^\text{shot}=1$ and $\alpha^\text{shot}_2=0$. 

The above modeling of the PS is very successful in the study of the broadband. However,  this is not useful for the Baryon Acoustic Oscillation (BAO) modeling because the damping effects due to long-wavelength displacements fields are not perturbative in a SPT scheme. To take account the BAO oscillations correctly, one employs infrared resummations (IR-resummations) \cite{Senatore:2014via}. The aim is to split the linear PS in a part without oscillations and other one with them, i.e., the wiggles removed piece, $P_{nw}$ and the wiggles piece, $P_{w}$. Then the linear PS is $P_L = P_{nw} + P_{w}$. Thus, the one-loop IR-resummed PS is \cite{Ivanov:2018gjr} 
\begin{align}\label{PsIR}
P_s^\text{IR}(k,\mu) &= 
 e^{-k^2 \Sigma^2_\text{tot}(k,\mu)} P_s(k,\mu) +  \big(1-e^{-k^2 \Sigma^2_\text{tot}(k,\mu)} \big) P_{s,nw}(k,\mu) \nonumber\\
 & \quad +  e^{-k^2 \Sigma^2_\text{tot}(k,\mu)} P_w(k) k^2 \Sigma^2_\text{tot}(k,\mu),
\end{align}
where
\begin{equation}\label{Sigma2T}
\Sigma^2_\text{tot}(k,\mu) = \big[1+ \mu^2 f \big( 2 + f \big) \big]\Sigma^2 + \mu^2 f^2 (\mu^2-1) \delta\Sigma^2,    
\end{equation}
and
\begin{align}
\Sigma^2 &= \frac{1}{6 \pi^2}\int_0^{k_s} dp \,P_{nw}(p) \left[ 1 - j_0\left(p \,\ell_\text{BAO}\right) + 2 j_2 \left(p \,\ell_\text{BAO}\right)\right],\\
\delta\Sigma^2 &= \frac{1}{2 \pi^2}\int_0^{k_s} dp \,P_{nw}(p)  j_2 \left(p \,\ell_\text{BAO}\right).
\end{align}
where  $j_n$ is the spherical Bessel function of order $n$. eq.(\ref{PsIR}) is the final expression that was used to compare simulated halo data in \cite{Aviles:2020wme}. In fact, the BAO scale is well determined and is given by $\ell_\text{BAO}\simeq 105 \hmpc$. Moreover, to fit data one uses the multipoles given by \cite{Rodriguez_Meza_2024}:
\begin{equation}\label{Pells}
P_\ell(k) = \frac{2 \ell + 1}{2} \int_{-1}^{1} d\mu \, P_s^\text{IR}(k,\mu) \mathcal{L}_{\ell}(\mu),    
\end{equation}
with $\mathcal{L}_{\ell}$ being the Legendre polynomial of degree $\ell$.

\subsection{fk-kernels approximation}
\label{fk-kernels}
In the previous subsections, the complete theory for calculating the 1-loop power spectrum is presented, using an IR-resummed scheme. However, the calculation of the 1-loop contributions depends on the kernels $F_n$ and $G_n$, which slow down the numerical computation. 

Furthermore, when using the MCMC method, the calculation is further delayed. To optimize this process, the fkPT approach \cite{Aviles:2021que} was developed, where an approximation of the kernels  is derived by considering only the $k$-dependence through the growth rate, that in turn is computed from eq. (\ref{f_MG}). This approximation is implemented in the codes, FOLPS-nu \footnote{https://github.com/henoriega/FOLPS-nu.git} \cite{Noriega:2022nhf} for neutrinos and fkpt\footnote{https://github.com/alejandroaviles/fkpt.git} \cite{Rodriguez_Meza_2024} for HS-$f(R)$. 

In the particular case of the full kernels for 1-loop contribution eqs.  (\ref{F2_kernel},\ref{G2_kernel}) and eqs. (\ref{F3kernel},\ref{G3kernel}) was demonstrated in \cite{Aviles:2021que} that we can evaluate them on the limit in large scales (LS). Then, the fk-kernels are defined by taking $k=0$, i.e.,
\begin{align}
F_2^{\texttt{fk}}(\vk_1,\vk_2) &=  F_2(\vk_1,\vk_2)\Big|_{\mathcal{A}=\mathcal{B}=\mathcal{A}^\text{LS}},   \\
G_2^{\texttt{fk}}(\vk_1,\vk_2) &=  G_2(\vk_1,\vk_2)\Big|_{\mathcal{A}=\mathcal{B}=\mathcal{B}^\text{LS}}, 
\end{align}
the values $\mathcal{A}^\text{LS}$, $\mathcal{B}^\text{LS}$ are obtained from eqs. (\ref{AandBdef}) for the second order kernels and
the same for $F_3$ and $G_3$ kernels with the corresponding $\mathcal{A}^\text{LS}$ and $\mathcal{B}^\text{LS}$ expressions. Therefore, using the fk-approach for the Symmetron, we only need to calculate the scale-dependent growth rate $f(k,t)$ and the limit of perturbative kernels at large scales. Also, using FFTLog method in FOLPS-nu one can speed up the calculation of loop integrals and hence, accelerate the numerical calculation in the analysis using an MCMC method.

The fkPT results are complemented with the  counterterm contributions that turn out to emulate missing terms from the full 1-loop kernels computations, as is demonstrated in \cite{Rodriguez_Meza_2024}. However, this approximation has to be tested model by model. For the Symmetron, we show in Figure \ref{fig:pls_fk_full}, the relative difference of the multipoles ($\ell_0$ in orange, and $\ell_2$ in blue) between full kernels and fk kernels for the Symmetron model. When considering the counterterm contribution proportional to $\alpha_0 k^2 P_L(k)$ of the form ($P_0 \rightarrow  P_0 + \alpha_0 k^2 P_L$) for the monopole and ($P_2 \rightarrow  P_2 + \alpha_2 k^2 P_L$) for the quadrupole in the fk-approximation (dashed lines), the differences lie within the 1 \%; for the quadrupole this is valid for  $k < 0.17 h  \text{Mpc}^{-1} $. This proves to be a good approximation; therefore, we conclude that the fk-approximation is viable for the Symmetron model and we will use it in the multipole calculations of next section. 

\begin{figure}
\begin{center}
    \includegraphics[width=4 in]{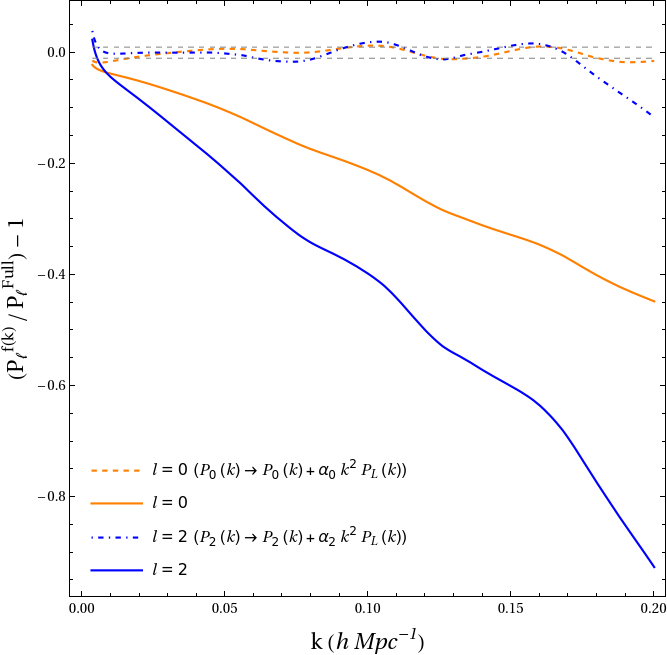}
    \caption{Relative difference between full kernels and fk kernels multipoles for Symmetron model with (dashed lines) and without adding counterterms (solid lines). }
    \label{fig:pls_fk_full}
\end{center}
\end{figure}

\end{section}
\begin{section}{ RSD multipoles for MG }\label{sect:multipoles}
In this Section, we compute the RSD multipoles for Symmetron model, comparing against $n=1$ HS-$f(R)$ and taking $\Lambda$CDM as a reference model. 
To compute the multipoles, we used Eqs. (\ref{PsIR}) and (\ref{Pells}) that take into account the full one-loop approximation and it is implemented in the code FOLPS-nu,  that is normally valid for $\Lambda$CDM plus neutrinos. In accordance, we implemented to it the two MG models studied here: Symmetron and a corrected version for the HS-$f(R)$. 

To obtain the linear PS for MG,  FOLPS-nu firstly employs the $\Lambda$CDM PS as input that is computed with CLASS \cite{Diego_Blas_2011} with a set of cosmological and nuisance parameters. Then, one employs the relation:
\begin{equation}
    P_L^{\text{MG}} (k)= \left( \frac{D_+(k)}{D_+(k \rightarrow 0)}\right)^2 P_L^{\text{$\Lambda$CDM}} (k),
    \label{PLMG}
\end{equation}
where $D_+(k)$ is the linear growth function for the MG model considered. Then, the full nonlinear theory of Sections \ref{sect:full 1-loop PS} and \ref{fk-kernels} is employed. 

To compute the  monopole and quadrupole we choose the same redshifts, at $z=0.5$ and $z=1$, and cosmological parameters as done in Ref. \cite{Aviles:2020wme}. There the multipoles were compared against ELEPHANT simulations for GR and F4, F5, and F6 models, using two halo mass catalogs ranges, \texttt{halos1}: $( 10^{12} < M_h < 4.5 \times 10^{12} M_\odot\ h^{-1} )$ and \texttt{halos2}: $(4.5 \times 10^{12} < M_h < 1 \times 10^{13} M_\odot\ h^{-1} )$ for both redshifts.
\begin{figure}[ht]
 	\begin{center}
  \includegraphics[width=3.02 in]{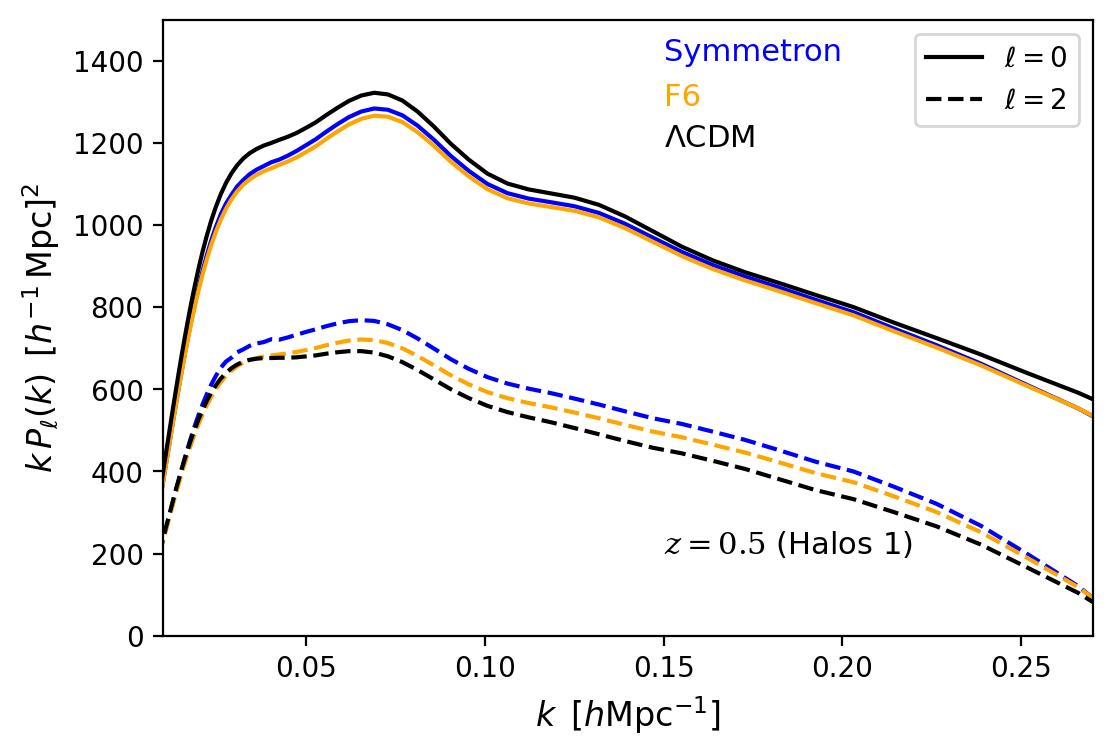}
  \includegraphics[width=3.02 in]{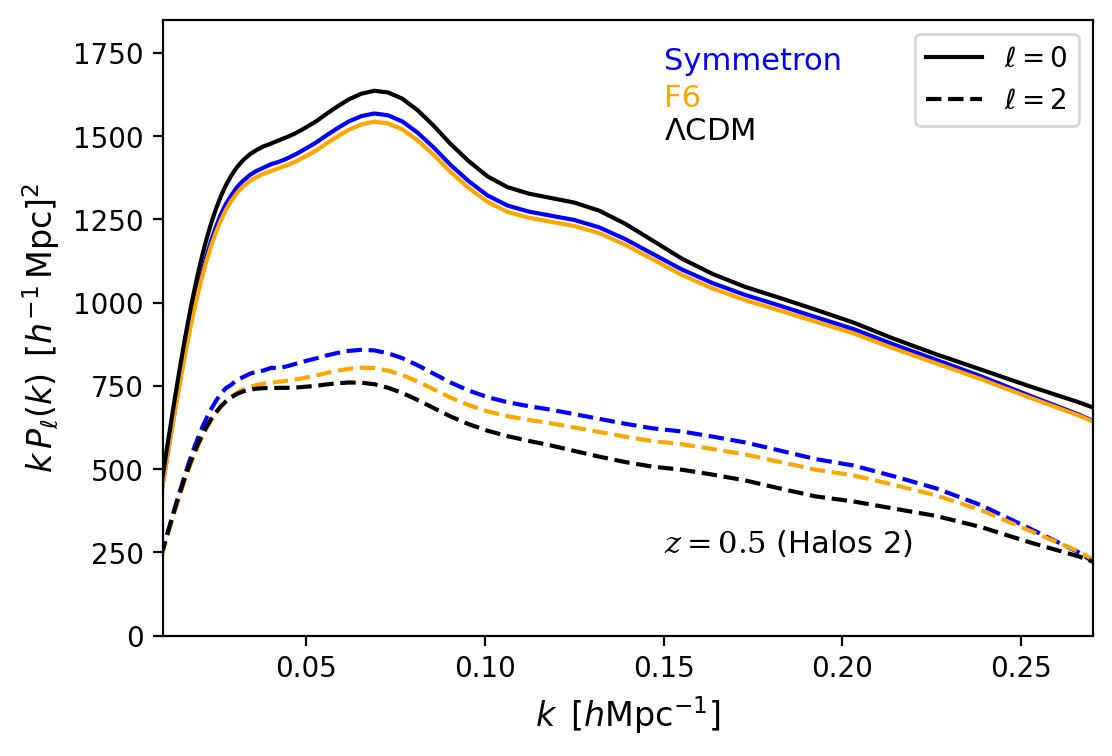}
  \includegraphics[width=3.02 in]{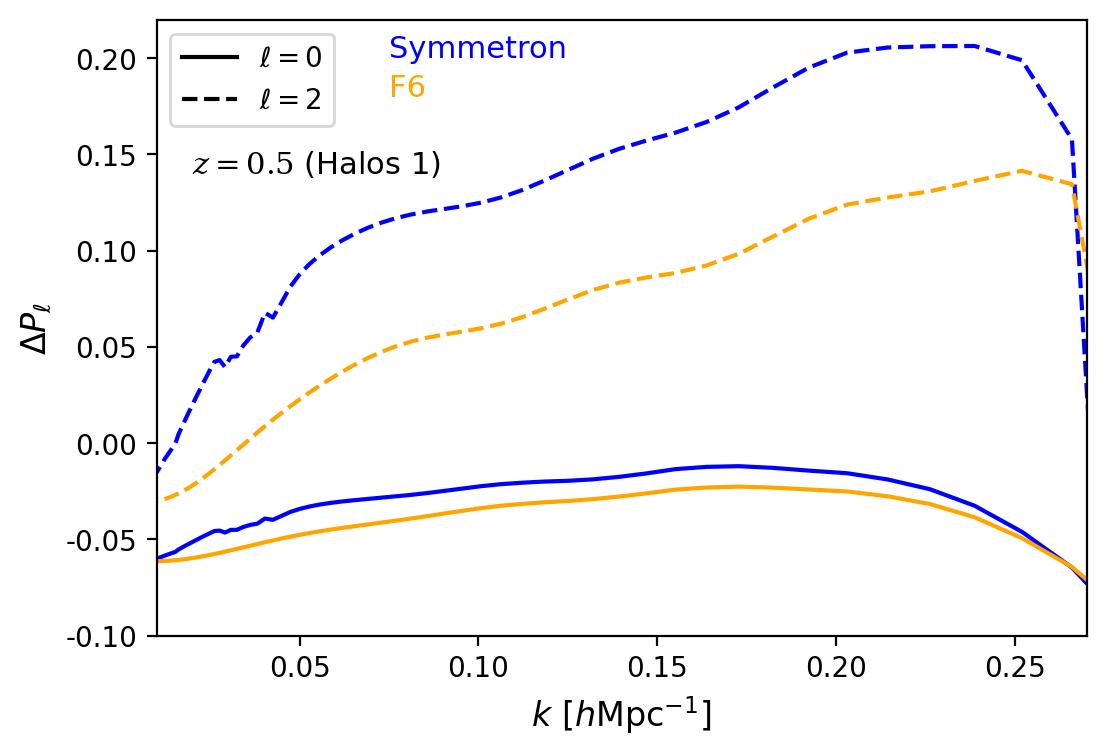}
  \includegraphics[width=3.02 in]{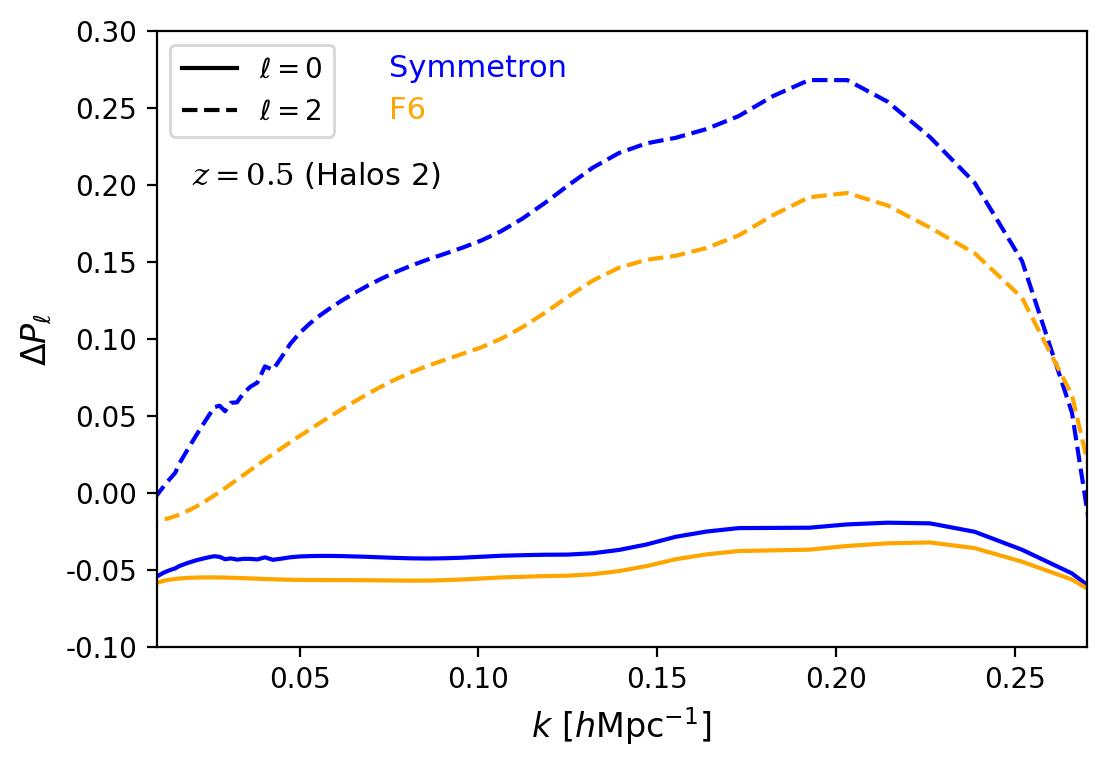}
 	\caption{Monopole ($\ell=0$) and quadrupole ($\ell=2$) at $z=0.5$ for \text{\texttt{halos1}} (left) and \text{\texttt{halos2}} (right) for the Symmetron, HS F6, and $\Lambda$CDM models. }
    \label{fig:Pl02h12z05}
 	\end{center}
\end{figure}
The cosmological parameters used are $\{ \Omega_m, \Omega_b, h, n_s, \sigma_8 \}=$ $\{ 0.281, 0.046, 0.697, 0.971, 0.848\}$ and nuisance parameters: four bias $\{ b_1, b_2, b_{s^2}, b_{3nl} \}$, four EFT counterterms $\{ \alpha_0, \alpha_2, \alpha_4, \tilde{c} \}$, and shot noise terms. As it is common, we use the co-evolution model \cite{Chan_2012,Baldauf_2012,Saito_2014} to reduce the independent parameters from 8 to 6,
\begin{eqnarray}
    b_{s^2} &=& -\frac{4}{7} (b_1 -1), \\
    b_{3nl} &=& \frac{32}{315} (b_1 -1),
    \label{coevolution}
\end{eqnarray}
where $b_1$ is the local bias. It is known that  MG models generate a smaller linear local bias than GR, $b_1^{\text{MG}} < b_1^{\text{GR}}$ \cite{Aviles:2018saf,Arnold:2018nmv}.  On the other hand, for the second order bias, $b_2^{\text{MG}} \geq b_2^{\text{GR}}$ as is shown in \cite{Valogiannis_2019} for ranges of \texttt{halos1,2}. 

Regarding the nuisance parameters, $b_1$ is the bias related to the linear part of the power spectrum and $b_2$ is a piece of the nonlinear part. Both parameters are used to adjust the broadband shape of the PS, with $b_1$ being the parameter that primarily induces a shift in amplitude when varied, while $b_2$ has a lesser contribution to this amplitude.  And since the linear part of Symmetron has the same scale dependence as HS-$f(R)$, we have decided to adopt the same bias parameter values as those of the F6 model. On the other hand, counterterms constants are unknown from the theory side, but they could be determined by fitting them to  simulations when available. Changing a counterterm constant affects the determination of the other counterterms when fitting the multipoles, so their values are somewhat degenerate.  This is true for both the F6 and Symmetron models. For simplicity, we decided to assume the same counterterm values for both models.

We show our results in  Figs. \ref{fig:Pl02h12z05} ($z=0.5$) and \ref{fig:Pl02h12z1} ($z=1.0$), where a black line corresponds to $\Lambda$CDM, an orange to F6 model, and a blue is for Symmetron. For these plots, we have set the three MG parameters of Symmetron, $\beta_0=1$, $m_0=1$, and $a_{ssb}=0.5$. As is mentioned above, this MG model only applies for $a > a_{ssb}$, and we fix the $\Lambda$CDM background behavior for values $a \leq a_{ssb}$. In both Figs. \ref{fig:Pl02h12z05} and \ref{fig:Pl02h12z1} we are considering the nuisance parameters of Table 1 of \cite{Aviles:2020wme} for GR and F6, and we use the same values as F6 for the Symmetron model because their results are quite similar, as earlier shown in Fig. \ref{fig:f_z} and now in Fig. \ref{fig:Pl02h12z05}. 
\begin{figure}
 \begin{center}
  \includegraphics[width=3.02 in]{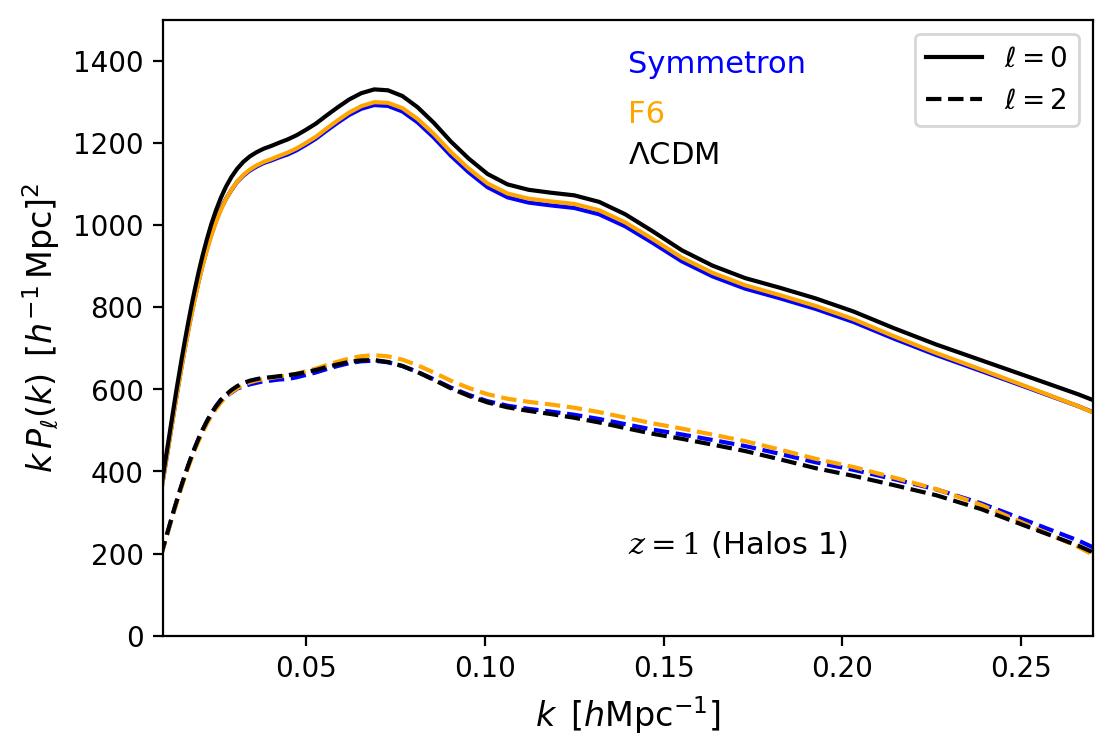}
  \includegraphics[width=3.02 in]{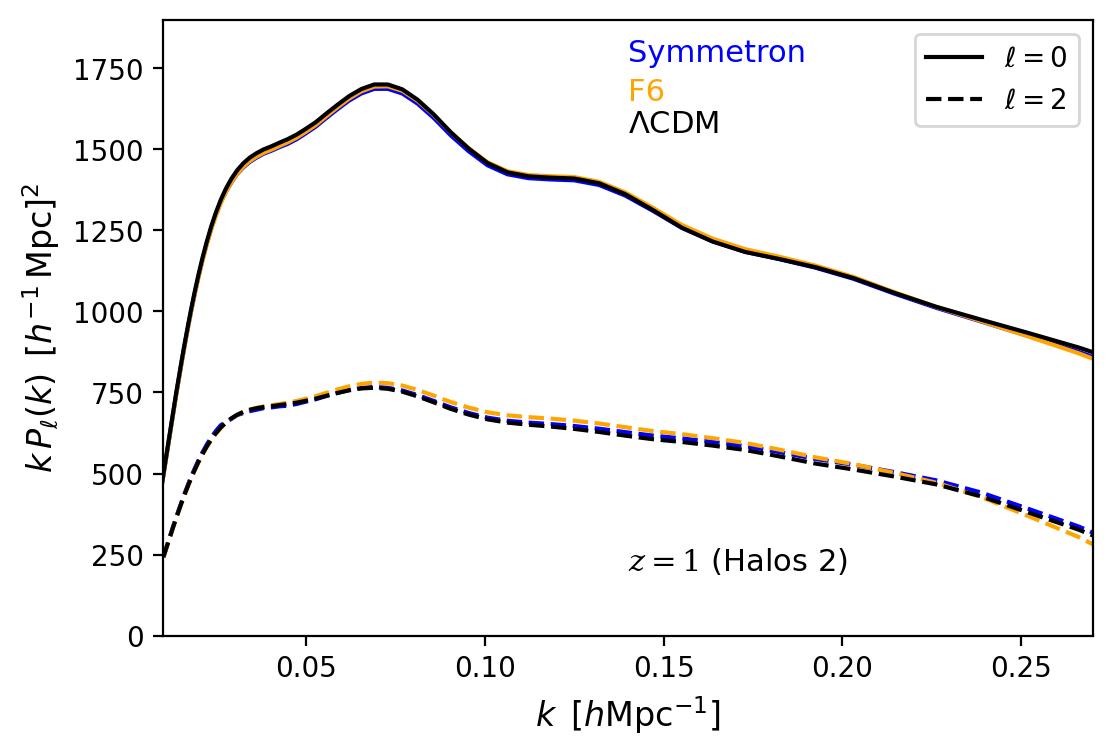}
  \includegraphics[width=3.02 in]{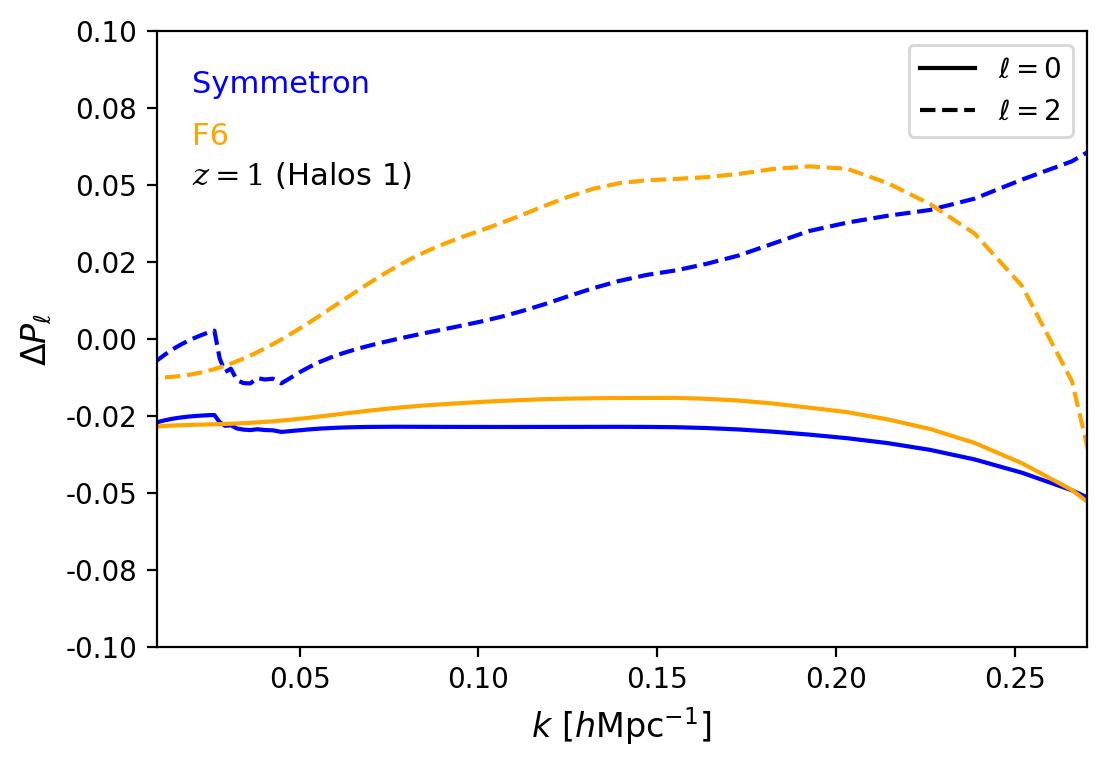}
  \includegraphics[width=3.02 in]{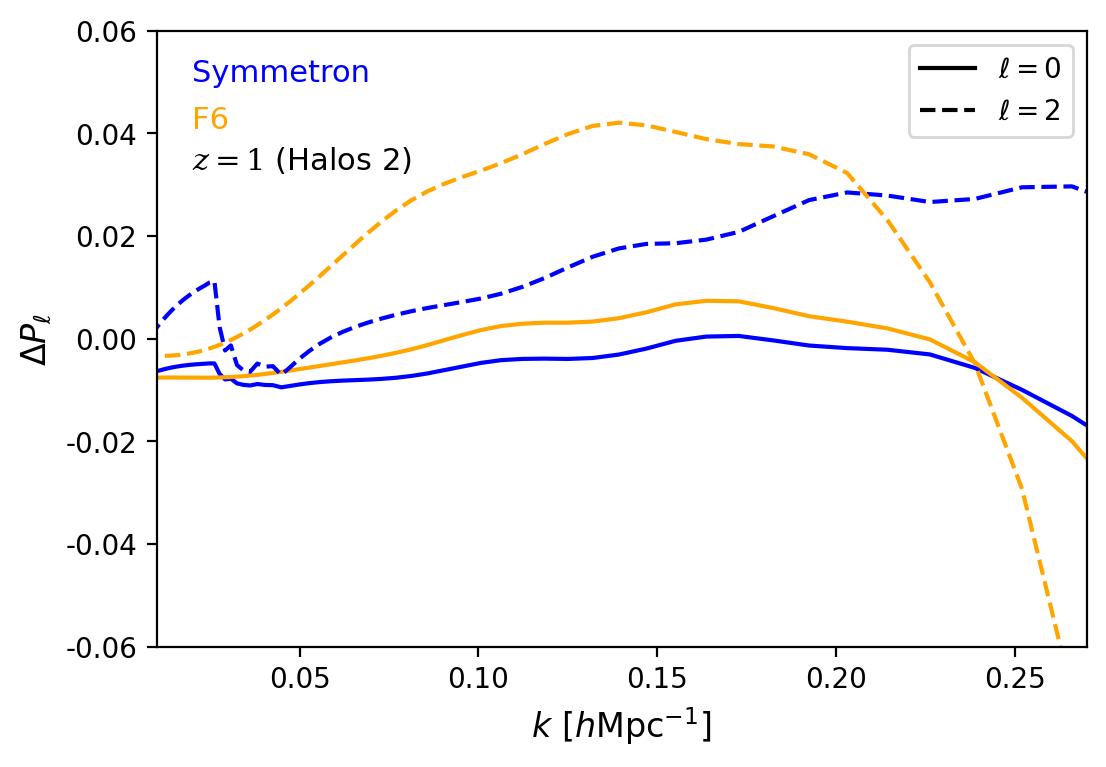}
 	\caption{ Monopole and quadrupole at $z=1$ from \text{\texttt{halos1}} (left) and \text{\texttt{halos2}} (right) for the Symmetron, HS F6, and $\Lambda$CDM models.}
    \label{fig:Pl02h12z1}
 	\end{center}
\end{figure}
We can see in Figs. \ref{fig:Pl02h12z05} and \ref{fig:Pl02h12z1}, the monopole for F6 and Symmetron is  smaller than the $\Lambda$CDM model in all cases, being Symmetron closer to $\Lambda$CDM at $z=0.5$ meanwhile F6 is closer to $\Lambda$CDM at $z=1$, for $k<0.25 ~h~ \text{Mpc}^{-1}$. The quadrupole has an opposite behavior, in all cases F6 and Symmetron are larger than $\Lambda$CDM for the two redshifts. These differences are better shown at $z=0.5$, as we can see in bottom panels of the two figures. 

Notice that we made a modification to eq. (4.61) of Ref. \cite{Aviles:2020wme} concerning to the IR-resummation scheme, when we implemented our models in FOLPS-nu, compared to the original version of the code, where the authors used  the approximation $f(k) \approx f_0$,  whereas we are taking the correct $k$-dependency $f(k)$. That change is responsible for the enhancement of the power for the computations using $f(k)$, as we can see in Figs. \ref{fig:f0vsfkz05} and \ref{fig:f0vsfkz05h2} in Appendix \ref{sect:appendix_B}. 

In the next subsection we fit the Symmetron using GR EZMocks data to test the recovery of the GR limit, that is,  when the coupling parameter, $\beta_0 \rightarrow 0 $, similar to Ref. \cite{Rodriguez_Meza_2024} for $f(R)$. 
\end{section}
\begin{subsection}{Model validation: EZMocks}
With the implementation of MG models in FOLPS-nu, using the fkPT approximation as we explained above, we want to make fittings with simulations. One might naively think of using some of the simulations of Symmetron model \cite{Ruan:2021wup}. However, current simulations for this model are not sufficiently precise to constrain the MG parameters. In \cite{Llinares_2014,Mirpoorian_2023}, the authors compared the quasi-static approximation (QSA) against the non quasi-static finding that QSA has implications on the matter distribution via the fifth force. In \cite{Ruan:2021wup}, three MG models are studied where Symmetron is one of them. In this work the aim is to construct an efficient and robust simulation code. Nevertheless, they only studied the linear matter PS and therefore that simulation is not suitable for testing the 1-loop RSD nor the IR-resummation scheme. For that reason we only tested our MG models with GR mocks, EZMocks \cite{BOSS:2016wmc}, to see if we recover GR from the Symmetron pipeline, as was done in \cite{Rodriguez_Meza_2024} for $f(R)$, where the authors used NSERIES GR simulations that consist of 7 realizations for $z=0.5$ and the same EZMocks for the computations of the covariances.

As part of the methodology, we need to calculate the covariance matrix. The cov matrix of the PS multipoles of one realization is computed by,
\begin{equation} \label{eq:cov_mat}
    C_{\ell \ell'} = \left\langle \left( \bar{P}_\ell(k_i) - \hat{P}_\ell(k_i) \right) \left( \bar{P}_{\ell'}(k_j) - \hat{P}_{\ell'}(k_j) \right)  \right\rangle,
\end{equation}
where the PS of a single realization at $k_i$ is $\hat{P}(k_i)$ and $ \bar{P}(k_i) =\langle \hat{P}(k_i) \rangle$ is the mean over the ensemble of realizations at $k_i$. 

For the covariance matrix construction, we use the 1000 EZMocks catalogs, also the covariance matrix is rescaled like $C_N = \frac{1}{N} C_{\ell \ell'}(k_i,k_j)$ with $N=7$, since we fit over the mean of 7 mocks. In addition, we apply the Hartlap factor \cite{hartlap2007your} to the cov matrix. 
\begin{figure}
 	\begin{center}
 	\includegraphics[width=3.5 in]{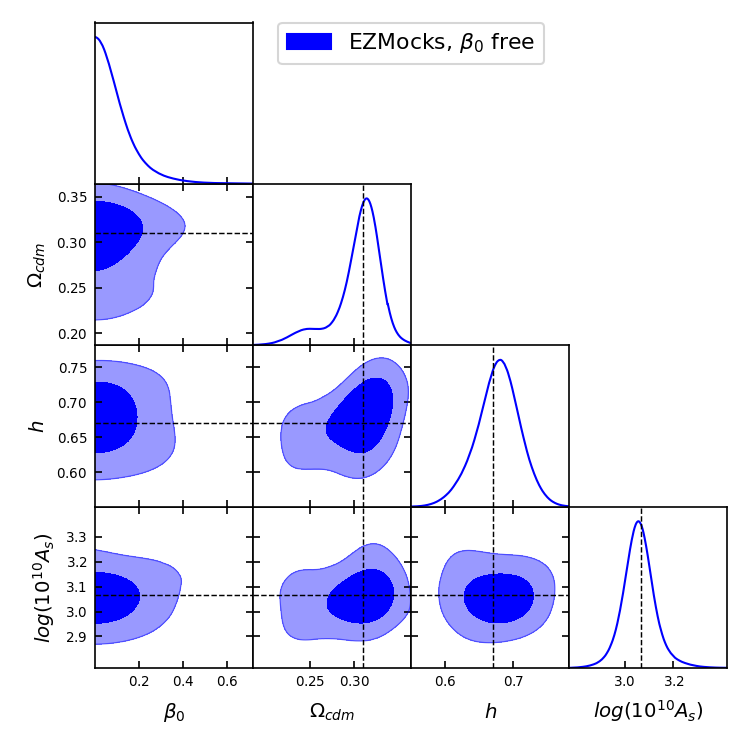}
 	\caption{Confidence level contours for parameter estimation performed on 7 EZMock realizations at $z=0.75$ for Symmetron model. The analysis employs the monopole and quadrupole PS multipoles, together with the covariance matrix estimated from 993 of the 1000 EZMock realizations. The coupling parameter $\beta_0$ is left free in the fitting with a flat prior $\mathcal{U}(-1.0,1.0)$. Dashed lines mark EZMocks parameter simulation values, and the best-fit values obtained are summarized in Table \ref{tab:PE_symm}. }
 	\label{fig:contour_symm}
 	\end{center}
\end{figure}
We take a maximum wavenumber value $k_{\text{max}}=0.17 ~h~ \text{Mpc}^{-1}$ in the PS because in Ref. \cite{Rodriguez_Meza_2024} was shown that it helps to estimate more correctly the $f_{R0}$ parameter. Besides, in Ref. \cite{eBOSS:2020hur} a test to recover cosmological parameters was done for fullshape analysis as a function of scale in each multipole of the PS, showing that the best $k_{\text{max}}$-value for the case monopole plus quadrupole is $k_{\text{max}}=0.15 ~h~ \text{Mpc}^{-1}$. Thus, our election is based on these arguments. The fiducial cosmology is $\{ \Omega_{cdm}, \Omega_{b}, h, \log(10^{10} A_s), n_s \} = \{0.31, 0.049, 0.67, 3.066, 0.97 \}$.

Fig. \ref{fig:contour_symm} shows the contours for the parameter estimation for the Symmetron model using EZMocks data. The aim was to recover the correct GR behavior from the Symmetron, allowing the MG parameter $\beta_0$ to vary. The same computations are done in Appendix \ref{sect:appendix_A} for the HS-$f(R)$ model. In both fittings, we used the same priors for cosmological and nuisance parameters, shown in Table \ref{tab:priors}. The Symmetron estimated parameters obtained are shown in Table \ref{tab:PE_symm}. Results show a good fit to the data, and recovering the expected $\Lambda$CDM results indicated with dashed lines in Fig. \ref{fig:contour_symm}.
\begin{table}[ht]
    \centering
    \begin{tabular}{c c}
    \hline
               & EZMocks, free $\beta_0$ \\ \hline \\
    $\beta_0$  & $< 2.8 \times 10^{-1}$   \\ \\
    $\Omega_m$ & $0.313^{+0.009}_{-0.025} $ \\  \\ 
    $h$        & $0.679^{+0.027}_{-0.032} $ \\ \\ 
    $\log(10^{10}A_s)$  & $3.05^{+0.052}_{-0.053} $  \\ \\
    $b_1$      & $1.698^{+0.281}_{-0.280} $ \\ \\ \hline \hline
    \end{tabular}
    \caption{Best fitted values for the MG parameter ($\beta_0$) at 95\% confidence level and the rest of the parameters  at 68\% confidence level.}
    \label{tab:PE_symm}
\end{table}

\end{subsection}
\begin{section}{Conclusions}\label{sect:conclusions}

The large-scale structure of the universe offers a unique arena for studying gravitational interactions through the growth rate and two-point statistics. It is of special interest nowadays when a hint of dark energy/modified gravity has been implied in the recent analysis of the DESI \cite{DESI:2024mwx,DESI:2024kob,DESI:2024aqx,DESI:2025zgx,
DESI:2025kuo, DESI:2025wyn} and DES \cite{DES:2025bxy} collaborations, among other analyses. In the present work, we discuss the observational consequences of two MG models: the HS-$f(R)$ model and the Symmetron, and compare our results with those of the standard $\Lambda$CDM model.  Results for the HS model have been previously reported in the literature; however, we computed them here to compare them directly with those of the Symmetron.  

At linear PT level, the gravitational modifications stem from two terms, the expansion rate and the source of the Poisson equation ($\mu(k,t)$ term), and both effects are encoded in the equation for the dark matter density contrast or in the growth rate, see Section \ref{sect:scale_dep_growth}. The growth rate for Symmetron, HS-$f(R)$, and $\Lambda$CDM models are presented in Figs. \ref{fig:f_z} and \ref{fig:f_k}, where in the former figure we plot it against redshift in two $k-$bins ($0.01 \leq k \leq 0.05$ and $0.06 \leq k \leq 0.1$ $h~\text{Mpc}^{-1}$) and in the later against the wavenumber for three redshits ($0.8, 0.5, 0$) to evaluate its effect after the Symmetron phase transition occurrence at $z_{\text{ssb}}=1.0$. From these plots we conclude that the Symmetron has a similar effect in the growth as  that of the HS F6 model.  The HS F4 and F5 models have a greater gravitational effect in the growth, but they are almost ruled out by observations \cite{He_2018}, thus Symmetron remains viable to be tested. Both MG models can be parameterized through an analytical function, eq. (\ref{fk_parameterization}), that turns out to be a good approximation for $k$-dependency, with an error less than $0.6\%$, cf. Fig. \ref{fig:fk_parameterization}. This formula could be of potential  interest when using an MCMC code to infer cosmological parameters, since its analyticity can save computational time. 

When we consider non-linear contributions in both models, HS-$f(R)$ and Symmetron, there is an important difference between them in the  scale dependence of $M_i$ functions that appear in the nonlinear term, $S(\Vec{k})$, in the Poisson equation (\ref{poisson}). Whereas for HS model $M_{2,3}$ are only time dependent, for the Symmetron both functions present $k$-dependence explicitly. Regardless of the specific $M_i$ functions that characterize the MG model, the formalism presented in Section \ref{sect:full 1-loop PS} remains applicable. However, the calculation of the 1-loop contributions requires numerical solutions of differential equations, which depend on the kernels $F_n$ and $G_n$. This process can significantly slow down the MCMC-based inference of cosmological parameters. To alleviate this delay, we adapted the fkPT approximation scheme to the Symmetron. However, this approximation needs to be tested, model by model, to guarantee that it is feasible when incorporating the EFT parameters; we did that and demonstrated that errors are less than 1\%.  Further, we corrected the growth function that enters in the resummation scheme of Section \ref{sect:full 1-loop PS}, correcting $f_0(t)$ for $f(k,t)$, as is meant in the resummation procedure for MG. Then, our results for the HS-$f(R)$ are new, and corrected those presented in Ref. \cite{Aviles:2020wme}.    

Next, we computed the RSD multipoles for the Symmetron, HS-$f(R)$, and $\Lambda$CDM models, and displayed our results in Figs. \ref{fig:Pl02h12z05} and \ref{fig:Pl02h12z1} 
for  redshifts $z=0.5$ and $1$. We find that the monopole and the quadrupole for both Symmetron and the HS F6 model are quite similar, corresponding to its similarity in their growth functions, discussed above. Finally, we continued to validate our pipeline with simulated data. However, we found no proper simulated data for the Symmetron model. Instead, we tested our pipeline with GR EZMocks data \cite{BOSS:2016wmc} to find out if Symmetron recovers GR results, i.e. as $\beta \rightarrow 0$, similar as it was done  \cite{Rodriguez_Meza_2024} for $f(R)$.  Our results show that indeed the GR parameters are obtained. At present, the fullshape covariances suitable for the DESI DR1 \cite{Zhao2024} or DR2 are not yet available to compute a proper parameter inference using real data, leaving this task for a future work. 

\end{section}

\acknowledgments

The authors thank Mario A. Rodr\'iguez-Meza and Alejandro Avil\'es for their valuable comments and  acknowledge support by SECIHTI (CONAHCyT) project CBF2023-2024-589. G. M-N. acknowledges support by SECIHTI (CONAHCYT) grant I01200/117/2024 and for the
PhD scholarship

\appendix
\begin{section}{ \texorpdfstring{Influence of $f_0$ and $f(k)$ in the multipoles}{Influence of f0 and fk in the multipoles} }\label{sect:appendix_B}
 The left panels of Fig. \ref{fig:f0vsfkz05} show the monopole and quadrupole for the \texttt{halos1} masses,  computed using $f_0$ in the IR resummation scheme given by eq. (\ref{PsIR}), and right panels show the same computation but using $f(k)$ at the same redshift $z=0.5$.  To obtain the results we employed the modified version of FOLPS-nu and considering the fkPT approach. In Fig. \ref{fig:f0vsfkz05h2} the same is computed for  the \texttt{halos2} masses. These plots refer to Figs. 7 and 8 in Ref. \cite{Aviles:2020wme} for HS-$f(R)$ in that case. Lower panels shown the relative difference between F6 and Symmetron model with respect to $\Lambda$CDM.  It is clear that the use of $f(k)$ causes an increase on the PS that is more prominent in the quadrupole. We have done the same calculations for $z=1$ and the same behavior is observed, but we decided to show only for $z=0.5$ for  comparison to results in Ref. \cite{Aviles:2020wme}. For this MG model, we set the three parameters $\beta_0=1$, $m_0=1$ and $z_{ssb}=1$. 
\begin{figure}
 	\begin{center}
  \includegraphics[width=3.02 in]{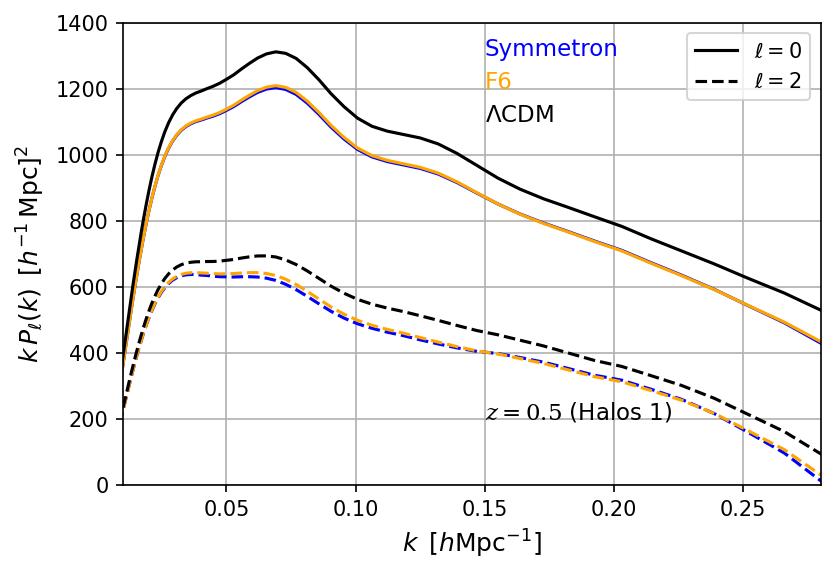}
  \includegraphics[width=3.02 in]{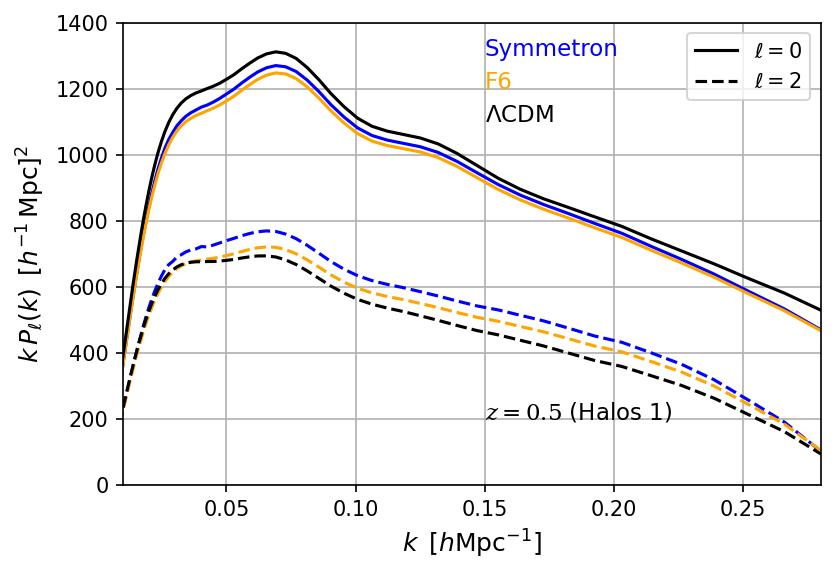}
  \includegraphics[width=3.02 in]{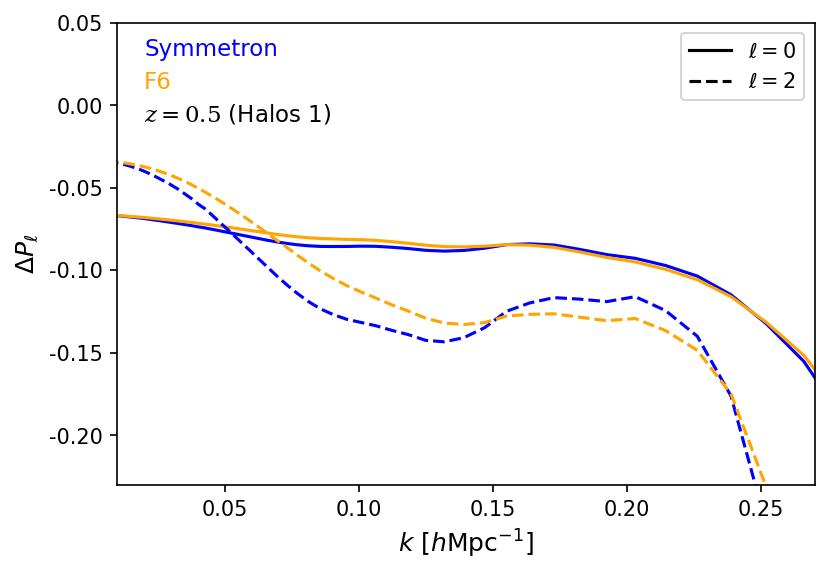}
  \includegraphics[width=3.02 in]{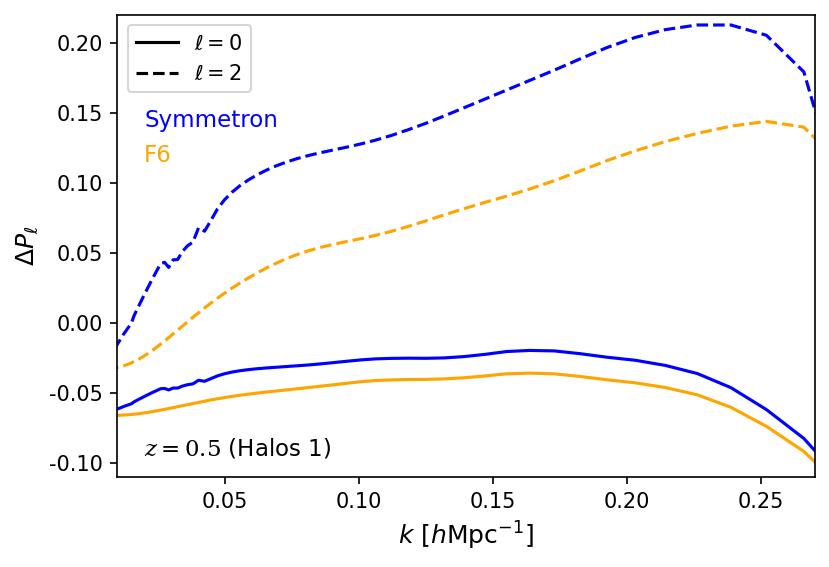}
 	\caption{Monopole and quadrupole at $z=0.5$ for \texttt{halos1} data  from Table 1 of \cite{Aviles:2020wme}. Left panel using $f_0$ and  right panel using $f(k)$. Both curves have the same values in all of the parameters. }
    \label{fig:f0vsfkz05}
 	\end{center}
\end{figure}
\begin{figure}
 	\begin{center}
  \includegraphics[width=3.02 in]{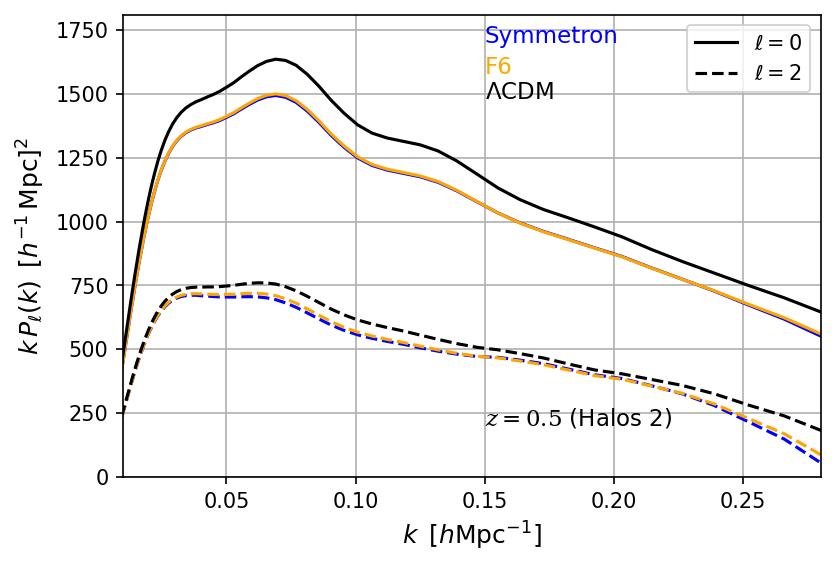}
  \includegraphics[width=3.02 in]{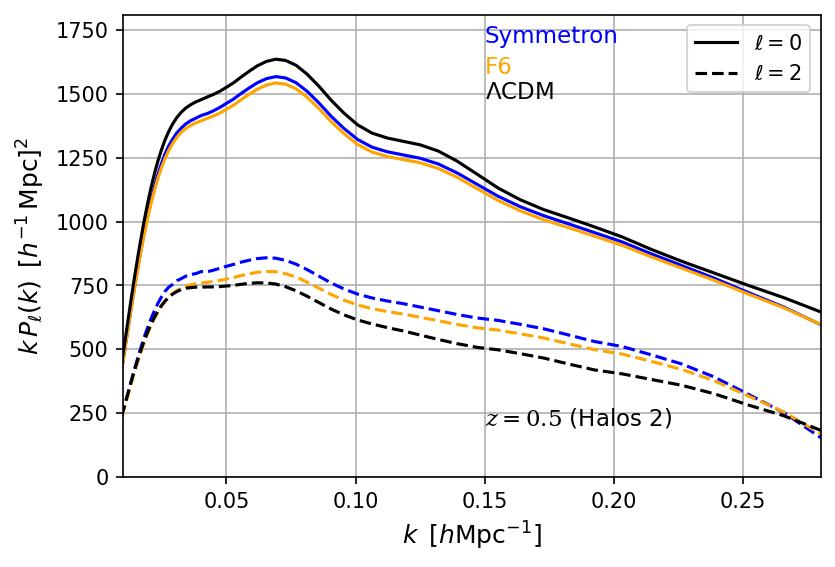}
  \includegraphics[width=3.02 in]{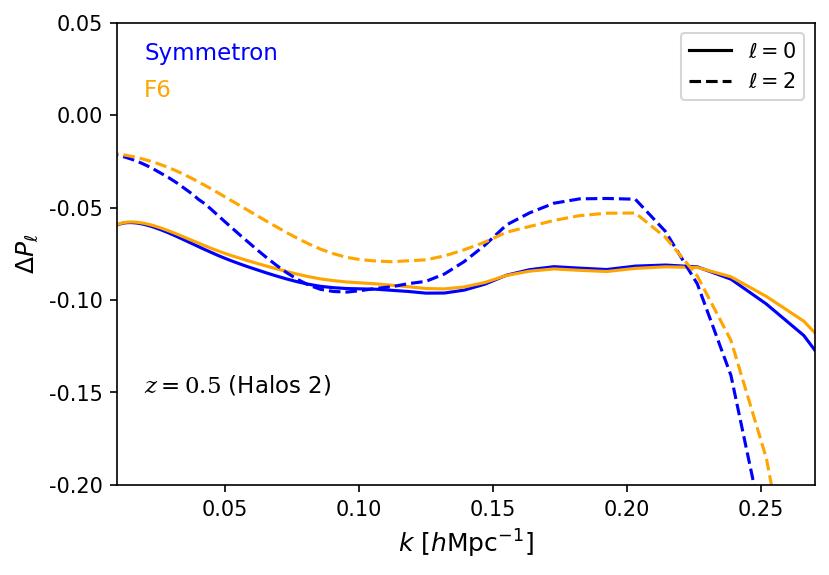}
  \includegraphics[width=3.02 in]{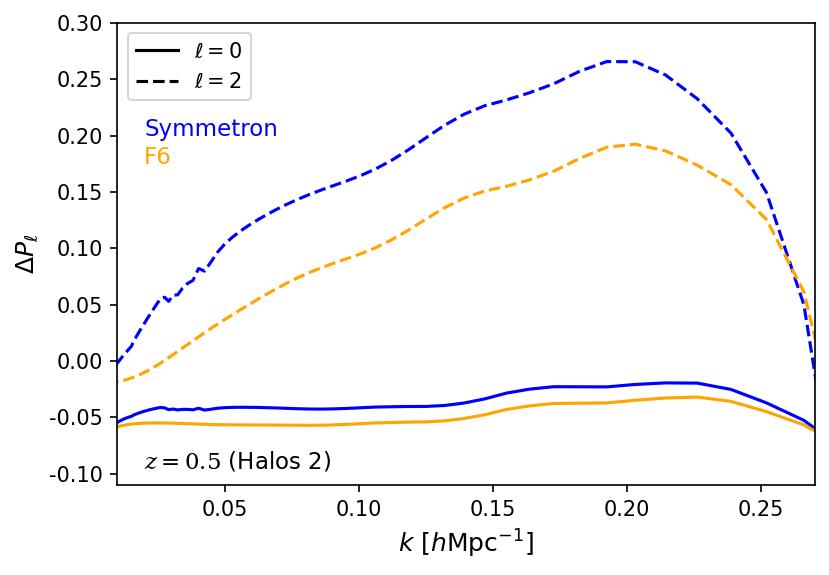}
 	\caption{Monopole and quadrupole at $z=0.5$ for \texttt{halos2} data  from Table 1 of \cite{Aviles:2020wme}. (Left) using $f_0$, (right) using $f(k)$. Both curves have the same values in all of the parameters. }
    \label{fig:f0vsfkz05h2}
 	\end{center}
\end{figure}

One may wonder whether the results yielded with $f(k)$ can be obtained when using $f_0$ with different nuisance parameters.  In Fig. \ref{fig:pls_fk_f0}, we are showing that it is not possible to obtain the same nuisance parameters, when using $f_0$ and $f(k)$, to  simultaneously adjust the monopole and quadrupole. This is because whereas the EFT parameters depend independently of each multipole and therefore can be adjusted, other parameters are the same for the different multipoles and cannot be changed for each multipole separately.
\begin{figure}
\begin{center}
    \includegraphics[width=3 in]{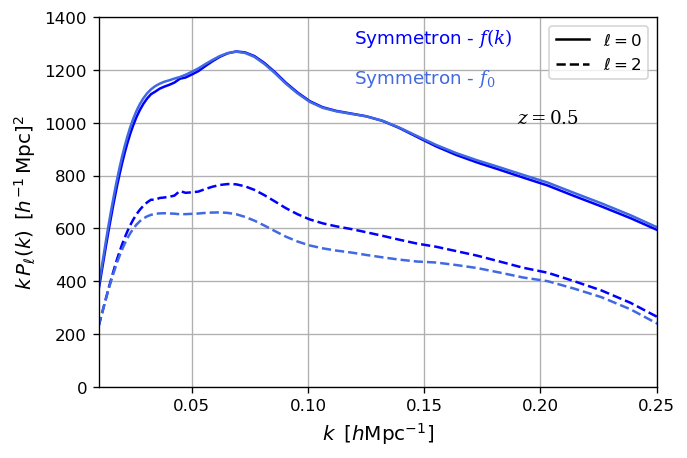}
    \includegraphics[width=3 in]{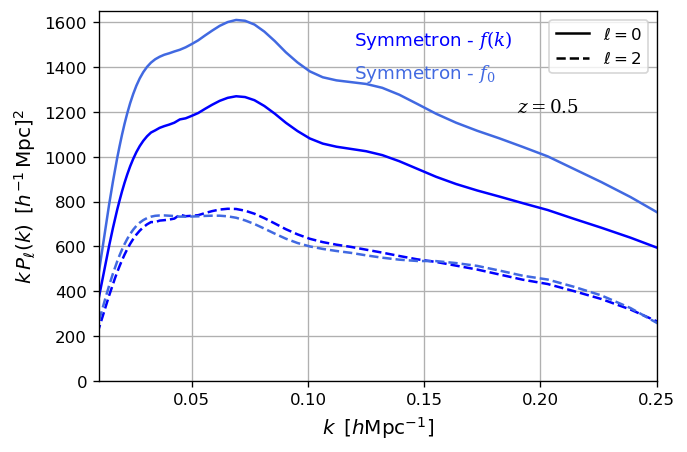}
    \caption{Multipoles behavior using $f(k)$ in the IR resumation with certain counterterm values, and results for an adjusted $f_0$ with other counterterms to reproduce the behavior of the monopole (left panel) and that of the quadrupole (right panel). The monopole and the quadrupole cannot be adjusted at the same time.}
    \label{fig:pls_fk_f0}
\end{center}
\end{figure}

\end{section}
\begin{section}{Validation of f(R)}\label{sect:appendix_A}
We present the validation of the modified version of code FOLPS-nu for the HS-$f(R)$ model, similar as in Ref. \cite{Rodriguez_Meza_2024}.
\begin{figure}[ht]
 	\begin{center}
 	\includegraphics[width=3.5 in]{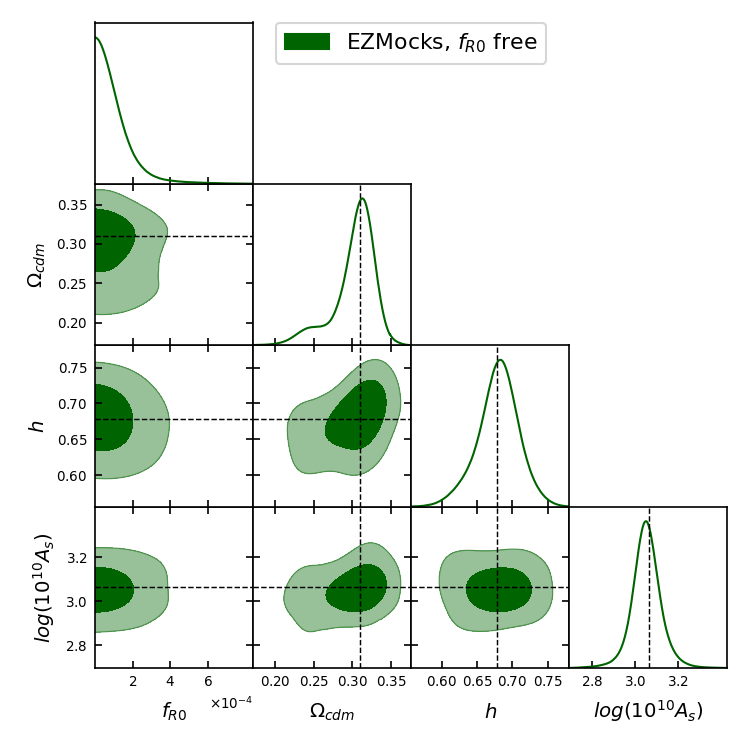}
 	\caption{ Confidence level contours for the parameter estimation performed on 7 EZMock realizations at $z=0.75$. The analysis employs the monopole and quadrupole power spectrum multipoles, together with the covariance matrix estimated from 993 of the 1000 EZMock realizations. The parameter $f_{R0}$ is left free in the fitting, and the priors adopted are listed in Table \ref{tab:priors}. The best-fit values obtained are summarized in Table \ref{tab:PE_fR}. }
 	\label{fig:fr_countours}
 	\end{center}
\end{figure}
Fig. \ref{fig:fr_countours} shows the confidence level contours for the cosmological parameters using the 1000 EZMocks \cite{BOSS:2016wmc}. These mocks are the same used in \cite{Rodriguez_Meza_2024}. Nevertheless, in contrast to \cite{Rodriguez_Meza_2024}, where the authors use the PS monopole and quadrupole data for 7 realizations of NSERIES cubic boxes; all these data are in this website\footnote{https://www.ub.edu/bispectrum/page12.html}. Here, we only use the 1000 EZMocks, since NSERIES and mock data have different redshifts. On the other hand, there is a difference in our IR-resummation scheme in comparison to ec. (3.70) of \cite{Rodriguez_Meza_2024}, where they consider $f_0$ instead $f(k)$ used here; the influence on the multipoles when choosing one or the other is shown in Figs. \ref{fig:f0vsfkz05} and \ref{fig:f0vsfkz05h2} of Appendix \ref{sect:appendix_B}. The fiducial cosmology is $\{ \Omega_{cdm}, \Omega_{b}, h, \log(10^{10} A_s), n_s \} = \{0.31, 0.049, 0.67, 3.066, 0.97 \}$. We use flat priors for all parameters except for baryon energy density, which we use a Gaussian prior from BBN constraint as it can be seen in Table \ref{tab:priors}. 

To compute the covariance matrix, eq. (\ref{eq:cov_mat}), we use the 1000 EZMocks and also we multiply the inverse covariance matrix by the Hartlap factor \cite{hartlap2007your}. Comparing our results against \cite{Rodriguez_Meza_2024}, the contours look a little different, but we can attach this behavior to the fact that we are doing the MCMC calculation only with EZMocks data since NSERIES GR simulation used in \cite{Rodriguez_Meza_2024} have different redshift. However, our result is consistent with the GR signal as we can see on the $68\%$ confidence level for all parameters shown in Table \ref{tab:PE_fR}.
\begin{table}
\begin{center}
\begin{tabular}{l c}
\hline
Parameter & Prior \\ \hline \\ 
$f_{R0}$ & $\mathcal{U}(-0.01,0.01)$ \\
$\Omega_{cdm}$ & $\mathcal{U}(0.1,0.4)$ \\
$\Omega_{b}$ & $\mathcal{G}(0.049,0.00001)$ \\
$h$ & $\mathcal{U}(0.4,0.9)$ \\
$\log(10^{10} A_s)$ & $\mathcal{U}(2,4)$ \\
$b_1$ & $\mathcal{U}(0.2,3.0)$ \\
$b_2$ & $\mathcal{U}(-10,10)$ \\
$\alpha_0$ & $\mathcal{U}(-200,200)$ \\
$\alpha_2$  & $\mathcal{U}(-200,200)$ \\
$\tilde{c}$ & $ \mathcal{U}(-1,1)$ \\
$\alpha^{\text{shot}}_0$  & $\mathcal{U}(0,50)$ \\
$\alpha^{\text{shot}}_2$  & $\mathcal{U}(-80,80)$ \\ \\ \hline \hline
\end{tabular} 
\caption{Priors used for each parameter in the MCMC computation. $\mathcal{U}$ is for flat priors and $\mathcal{G}$ is for gaussian prior.}
\label{tab:priors}
\end{center}
\end{table}
\begin{table}
    \centering
    \begin{tabular}{c c}
    \hline
               & EZMocks, free $f_{R0}$ \\ \hline \\
    $|f_{R0}|$  & $< 3 \times 10^{-4}$   \\ \\
    $\Omega_m$ & $0.313 ^{+0.008}_{-0.028}$  \\ \\ 
    $h$        & $0.681 ^{+0.023}_{-0.026}$  \\ \\ 
    $\log(10^{10}A_s)$  & $3.053 ^{+0.050}_{-0.049}$  \\ \\
    $b_1$      & $1.668 ^{+0.291}_{-0.265}$  \\ \\ \hline \hline
    \end{tabular}
    \caption{Best fitted values for the $f_{R0}$ parameter at 95\% confidence level, cosmological parameters and local bias $b_1$ at 68\% confidence level. These results are computed at $z=0.75$ for the mean PS of 7 mocks of 1000 EZMocks.}
    \label{tab:PE_fR}
\end{table}
\end{section}
\newpage
 \bibliographystyle{JHEP}  
 \bibliography{growth_MG.bib}  

\end{document}